\documentstyle[preprint,prb,aps]{revtex}
\begin{document}
\draft
\input{psfig}
\title{DOMAIN WALLS IN THE QUANTUM TRANSVERSE ISING MODEL}
\author{Malte Henkel,$^1$ A. Brooks Harris,$^{1,2}$
and Marek Cieplak$^3$}
\address{ (1) Theoretical Physics, Oxford University,
1 Keble Rd. Oxford OX1 3NP, UK}
\address{(2) Department of Physics,
University of Pennsylvania, Philadelphia, PA 19104-6396}
\address{(3) Institute of Physics, Polish Academy
of Sciences, 02-668 Warsaw, Poland}
\date{\today}
\maketitle
\begin{abstract}
We discuss several problems concerning domain walls in the
spin $S$ Ising model
at zero temperature
in a magnetic field,
$H/(2S)$,
applied in the $x$ direction.
Some results are also given for the planar ($y$-$z$) model in a
transverse field.  We treat the quantum problem
in one dimension by perturbation theory at small $H$ and
numerically over a large range of $H$.  We obtain the spin
density profile by fixing the spins at opposite ends
of the chain to have opposite signs of $S_z$. One
dimension is special in that there the quantum width of the
wall is proportional to the size $L$ of the system.  We also
study the quantitative features of the `particle' band which
extends up to energies of order $H$ above the ground state.
Except for the planar limit, this particle band is well separated
from excitations having energy $J/S$ involving creation of more
walls.  At large $S$ this particle band develops energy gaps
and the lowest sub-band has tunnel splittings of order
$H2^{1-2S}$.  This scale of energy gives rise to anomalous
scaling with respect to a) finite size, b) temperature, or
c) random potentials.  The intrinsic width of the domain wall
and the pinning energy are also defined and calculated in
certain limiting cases.  The general conclusion is that quantum
effects prevent the wall
from being sharp and in higher dimension would
prevent sudden excursions in the configuration of the wall.
\end{abstract}

\pacs{PACS numbers: 75.30.Et, 71.70.Ej, 75.30.Gw, 75.10.N, 75.10.H}

\section{INTRODUCTION}

Recently, there has been a growing interest in the study
of interfaces with non-trivial geometry. Such interfaces arise
in a variety of situations including domain walls in random
magnets,\cite{FAMILY,HUSEH,AJBMAM,MCJRB,MCLI,CMB} fluid invasion
in porous media\cite{FAMILY,MCMR}, spreading on heterogeneous
surfaces\cite{MORJFJ}, membranes and vesicles in
biology,\cite{LIP} and epitaxial growth in materials
science.\cite{FAMILY,THORN}

In connection with such problems it is natural to ask whether
quantum effects play a significant role.  For static properties
it is well established that in nonrandom systems, such as a
spin $S$ antiferromagnet, with only a nearest neighbor exchange
interaction, $J$, there are various regimes.  Near the critical
temperature at $T_N \sim JS^2$, thermal fluctuations are dominant.
In the ordered phase, as long as $T_c/S \ll T \ll T_c$,
quantum effects due to the finiteness of $S$ are unimportant.
For $T< T_c/S$ one is in a quantum regime, where the quantum
statistics of spin waves and their interactions leads to
dependences on $S$
and $T$ not present in the classical
($S \rightarrow \infty$) limit.  Heuristic arguments indicate that
quantum effects could influence the nature of domain walls in spin
systems.  In the picture in which the trajectory of the
domain wall is likened to the trajectory of a particle in space
as a function of time, quantum effects cause a
smearing out of the trajectory.
For such a system analytic
and numerical
work is obviously
very difficult.  Accordingly, we have been led to carry
out a program of analytic work for quantum domain walls
in one dimension.  For this purpose we consider domain
walls at zero temperature $T$
in the spin $S$ Ising model (with nearest-neighbor
exchange coupling $J$) in a transverse field, $H$,
whose Hamiltonian is
\begin{equation}
{\cal H}_{TI} = - {H \over 2S} \sum_n S_x(n) -
{J \over 2S^2} \sum_n S_z(n) S_z(n+1) \ ,
\end{equation}
where ${\bf S}(n)$ is a quantum spin $S$ operator at site $n$.

To study domain walls in this model we introduce boundary
conditions in which the spin at one end of the chain is
fixed to be ``up'' and that at the other end is fixed to be
``down.''  This model has some interest in its own right.
For $S=1/2$ its properties can, in principle, be related to
those of the associated free Fermion model obtained via the
Jordan-Wigner\cite{Kats62}
transformation.  However, with domain-wall boundary
conditions, this relation is not easy to implement.
Accordingly, we approach the analysis of the properties
of domain walls in this model via perturbation theory
for $H/J \ll 1$ and more generally via numerical
solution for the wave functions and energies of the
ground state and the low-lying excited states.
Needless to say, some of the properties of this model in
one dimension can not be extrapolated to higher
dimensional systems.  However, in most cases, one
can safely say which features can be so extrapolated
and which can not.

We may summarize our results for the one dimensional model
as follows.  The magnetization profile has a width
of order the length of the chain.  The low-lying excited
states
comprise a
manifold of ``particle'' states, which
results
when the center of the wall propagates from
site to site. These results are easily understood within
perturbation theory in $H/J$.  In the classical limit the
width of the wall is of order the correlation length,
i. e. it is of order a lattice constant, as long as
one is far from the critical regime
at $H\approx 2J$, above which long range order disappears.
In the classical limit and for small $H/J$, we evaluate
a barrier energy, analogous to the Peierls-Nabarro
energy,\cite{PEIERLS} which prevents the free motion of the
domain wall.  When the correlation length becomes very
large this barrier energy becomes small and since it
is harder to calculate in this limit, we did not
attempt such calculations.  It would be interesting
to calculate this energy for a quantum system, but
in the present case, since the width of the wall is of
order $L$, we can say that this barrier energy vanishes
in the thermodynamic limit.
Also, in the large $S$ limit, we find that the
quantum hopping of the domain wall from one site to
the next is actually analogous to a tunneling process,
so that the hopping matrix element is not of order
$H$,
as it is for the low spin case, but is now of order
$H \exp (-aS)$, where $a=\ln 2$.  It would be of some
interest to recover this result within a field theory
of one space and one time dimension to describe such
a quantum effect.

Briefly, this paper is organized as follows.  In Sec. II
we define the class of models we will analyze in one
dimension.  In Sec. III we give numerical results for
the magnetization profile for $S=1/2$, $S=1$ and $S=3/2$
and compare these to analytic
results we obtain using perturbation theory in the small
$H/J$ limit.  In Sec. IV we give analytic results for large $S$.
The classical results for $S=\infty$ are given both in the
continuum limit (i.e., when the wall is very broad) and in the
discrete limit (when the wall is very narrow).  Here we also
analyze the quantum system for large $S$.  In Sec. V we give
various results concerning the nature of the energy spectrum in
the presence of a domain wall.  We give numerical and analytic
results for the nature of the particle spectrum caused
by the matrix element which allows the wall to hop
from one site to the next.
Finally, in Sec. VI
we summarize the conclusions to be drawn from our work.

\section{DEFINITION OF ONE DIMENSIONAL MODELS}

The first one-dimensional model we consider is the spin $S$ Ising
model in a transverse field, with the Hamiltonian
\begin{equation}
\label{HAMZZ}
{\cal H}_{\rm TI}
= - {H \over 2S} \sum_{n=1}^L S_x (n) -
{J \over 2S^2} \sum_{n=0}^{L} S_z(n) S_z(n+1) \ .
\end{equation}
This formulation of the model has the advantages that
a) the domain wall energy (for $H=0$) is independent
of the value of $S$ and is equal to $J$, and b) the
mean-field transition temperature $T_0$
(at which significant correlations begin to develop between
neighboring spins) is of order $J$ independent of $S$.
To discuss domain walls, the spins $S(0)$ and $S(L+1)$ will
be fixed by boundary conditions, as discussed below.
Thus $L$ is the number of ``active'' spins in the chain.
We will often use the notation $h=H/J$.

We will also consider the ``yz" model in a transverse field,
for which the Hamiltonian is
\begin{equation}
\label{YZEQ}
{\cal H}_{\rm yz} = - {H \over 2S} \sum_{n=1}^L S_x (n) -
{J \over 2S^2} \sum_{n=0}^{L} \Biggl(
S_z(n) S_z(n+1) + \epsilon  S_y(n) S_y(n+1) \Biggr) \ .
\end{equation}
The most important difference between these two models is
that the interaction term (proportional to $\epsilon$) in
${\cal H}_{\rm yz}$ allows the spins to tip away from the
$z$-axis.  For $H=0$ and with free-end boundary conditions,
the ``y-z" model with $\epsilon=1$ has the
continuous $U(1)$ symmetry instead of the discrete
$Z_2$ symmetry of the Ising model which results when
$\epsilon\not= 1$.  In fact, we recall
the ground state phase diagram\cite{Kats62} of this more
general model, shown in Fig. 1 for spin $S=\frac{1}{2}$,
where one sees (at zero temperature) the disordered phase (D),
the ordered ferromagnetic Ising phase (F), and the ordered
oscillatory phase (O). In both ordered phases the spontaneous
magnetization $\langle S_z(n) \rangle$ is non-vanishing
(where $\langle \cdots \rangle$ denotes the thermodynamic
average at temperature $T$). However, the (connected) two-point
correlation function, $G(R)= \langle S_z(R) S_z(0) \rangle
- \langle S_z(R) \rangle \langle S_z(0)\rangle$,
behaves as follows\cite{Kats62} for $R\rightarrow\infty$:
\begin{equation} \label{Corrs}
G(R) \sim \left\{
\begin{array}{cl}
R^{-1/2} \exp\left( - R/\xi \right) , & \mbox{\rm  phase D} ; \\
R^{-2} \exp\left( -R/\xi \right)  ,   & \mbox{\rm  phase F} ; \\
R^{-2} \exp\left( -2R /\xi \right) \cdot {\rm Re} B e^{i K R} , &
\mbox{\rm  phase O} ;
\end{array} \right.
\end{equation}
where the correlation length $\xi$ is
a known function of $h$ and $\epsilon$
in each of the three phases,
$B$ is a constant and $\cos K = \sqrt{ h^2 /(4 \epsilon)}$.
When going from phase F to phase O, there is a new diverging length
$\Lambda \sim 1/K$, which is not simply related to a gap in the
spectrum of ${\cal H}_{\rm yz}$ (as is $\xi$),
but indicates the wavelength at which the correlation function
at large distances oscillates under an exponentially decaying
envelope.

To discuss the width and energy of a domain wall, we shall
work with ``up-down'' boundary conditions, in which
we require
\begin{equation}
\label{BC}
S_z(0)=-S_z(L+1)=S \ .
\end{equation}
We can then study the profile of the wall by evaluating

\begin{equation}
M(n) = \langle 0 | S_z(n) | 0 \rangle /S \ ,
\end{equation}
where $|0\rangle$ denotes the ground state. As we discuss in more
detail below, in order to obtain a spin
profile for a {\em quantum} system, antiperiodic boundary conditions
which introduce a boundary coupling $(K/2) S_z(0) S_z(L+1)$
{\em cannot} be used in a naive way.  However, for a
classical system such antiperiodic boundary conditions will
prove convenient.  The energy of the domain wall, $E_w$, is defined to
be $E_w=E-E_p$, where $E$ is the energy with ``up-down'' or
antiperiodic boundary conditions and $E_p$ that with ``up-up''
or periodic boundary conditions.

\section{RESULTS FOR THE PROFILE FUNCTION FOR $S=1/2$, $1$ and $3/2$}

\subsection{Numerical Results for the Ferromagnetic Phase}

We begin by presenting numerical results for the profile function
$M(n)$ for the ferromagnetic phase F.
For general spin we are constrained to
rather small systems, since the number of quantum states is
$(2S+1)^L$ for $L$ sites (excluding the fixed boundary spins).
So why not use translational symmetry to reduce the number of states?

Indeed, that would be possible for the special case of
antiperiodic boundary conditions (with $K=J$)
mentioned in the previous section. These boundary conditions are
peculiar since $\cal H$ commutes with the modified translation
operator $\widetilde{{\cal T}}=\sigma_x(0) {\cal T}$ where
$\cal T$ is the usual translation operator and $\sigma_x(0)$ changes
the sign of $\sigma_z(0)$. In addition, $\cal H$ can be decomposed
(for any $K$) into a block diagonal form. The corresponding states
are said to be in the even or odd sector, respectively, depending on
their parity under spin reversal. Furthermore, $M(n)=0$ in that
case and one should consider instead the quantity
$\widetilde{M}(n)=\langle 0| S_z(n)|1\rangle$ where $|0\rangle$ and
$|1\rangle$ are the ground states in the even and odd sectors
(with respect to the Ising symmetry) of the model, respectively.
It is well-known that at least for {\it periodic}
boundary conditions this defines\cite{Uzel81} a sensible
finite-lattice approximation to the order parameter (see  Ref.
\onlinecite{Chri93} and references therein).  However, for $K=J$,
even for $L$ finite, there are two {\em degenerate} states
$|1\rangle$, $|1'\rangle$ in the odd sector,
one of which gives rise to the desired
wall profile, while the profile of the other one is essentially flat.
On the other hand, for $K > J$, one does get a wall profile, but
its shape depends on the value of $K/J$.  We avoid these
problems by choosing the ``up-down'' boundary conditions
defined above in Eq. (\ref{BC}). Furthermore,
since we are interested in situations
far from criticality, the convergence of the finite-size data with
$L\rightarrow\infty$ is usually quite rapid so that the moderate sizes
achieved are sufficient for our purposes.

The determination of the lowest eigenstates of the Hamiltonians
${\cal H}_{\rm TI}$ and ${\cal H}_{\rm yz}$ using the Lanczos
algorithm is fairly standard\cite{Chri93}. From the ground state
$|0\rangle$ we calculate the local magnetization $m(r)$
\begin{equation} \label{MagPro}
m(r) = \langle 0| S_z (n)/S |0\rangle \;\; , \;\; r = n/(L+1) \ .
\end{equation}

Our results pertaining to the ferromagnetic phase F are as
follows. In Fig. 2, we display $m(r)$ for
$h=0.1$ and $\epsilon=0$ and $S=\frac{1}{2}$ for various values of
$L$. With the exception of the smallest $L$-values used,
we first observe that the data collapse onto a single curve,
which implies that the $L$ considered are
already sufficiently large as compared to the correlation length
that they represent the $L \rightarrow \infty$ limit.
Second, we note that the profile is quite wide and a continuous
function of $r$. We find a similar data
collapse for all the situations
we are going to consider in the sequel.
Next, we show in Fig. 3, that the overall shape of the
profile is not a peculiarity of having spin $S=\frac{1}{2}$. Rather,
we see that the profiles obtained with the parameters $\epsilon=0$ and
$h=0.1$ (Fig. 3A) and $h=0.5$ (Fig. 3B)
both spins $S=1$ and $S=\frac{3}{2}$ are very close to the one for
$S=\frac{1}{2}$. Nevertheless, the finer details of the respective
shapes become different as $H$ increases.
Finally, in Fig. 4, we show the effect of varying
$h$ in the $S=\frac{1}{2}$ case. Note that the profiles for
a transverse field as small as $h=10^{-5}$ and as large as $h=0.1$
are superimposed onto each other. The fact that the
width $w$ of the domain wall is proportional to the size of the
system is a peculiarity of a one dimensional quantum model.
In higher dimensions we would expect $w$ to remain finite
as $L \rightarrow \infty$ except possibly at a critical
point where the correlation length diverges.

We shall turn to a quantitative explanation of these findings below.
Profiles for the oscillatory phase are discussed in subsection C.

\subsection{Simplified Calculations for the Small $S$ Quantum Case}

In order to gain some understanding
of the results of the last section,
we now present some simple approximate calculations. As it turns out,
most of the physics of the problem for $h$ small is conveniently
described in terms of degenerate perturbation theory involving the
manifold $\cal M$ of the $h=0$ ground states. This procedure avoids
the additional technical complications of the free fermion method for
``up-down'' boundary conditions.
The boundary conditions are that spin $S(0)$ is fixed to have $S_z=S$
and spin $S(L+1)$ has $S_z=-S$.  For $S=1/2$ the manifold ${\cal M}$
contains the states $|n\rangle$, for $n=1, 2,\dots L+1$, where
$|n \rangle$ denotes the quantum state (shown in Fig. 5)
in which spins $i$ with $i < n$ have their
$z$-component of spin equal to $+1/2$ and those with $i \geq n$ have
their $z$-component of spin equal to $-1/2$.
The only nonzero matrix elements of the
Hamiltonian within the ground manifold are
\begin{eqnarray}
{\cal H}_{n,n}  & = & E_0 \equiv - {1 \over 2} (L-1)J \ , \ \
1\leq n \leq L+1 ; \nonumber \\
{\cal H}_{n,n+1} & = & {\cal H}_{n+1,n} = {1 \over 2} H \ , \ \ \ \ \
1 \leq n \leq L \ .
\end{eqnarray}
Considering only the manifold ${\cal M}$, one finds the corresponding
eigenvectors $\psi_p$, and eigenenergies
$E_p$ (for $p=1,2, \dots L+1$)
to be
\begin{equation} \label{PertTh}
\psi_p = C \sum_{n=1}^{L+1} \sin[np\pi/(L+2)] |n \rangle \ , \ \ \ \
E_p = E_0 + H \cos [p \pi / (L+2)] \ ,
\end{equation}
where $C$ is a normalization constant.
Perturbative corrections to the energy will
occur at
order $H^2/J$ and to the wavefunction at order $H/J$.  We
may calculate the profile function within the small $h$
approximation:
\begin{eqnarray}
M_p(n) \equiv \psi_p^\dagger ( S_{nz} /S)
\psi_p & = &  { \sum_{k=n+1}^{L+1}
\sin^2 [kp \pi / (L+2)] -
\sum_{k=1}^n \sin^2 [kp \pi / (L+2)] \over
\sum_{k=1}^{L+1} \sin^2 [kp \pi / (L+2)] } \nonumber \\
&=& 1 - 2 { \sum_{k=1}^n \sin^2 [kp \pi / (L+2)] \over
\sum_{k=1}^{L+1} \sin^2 [kp \pi / (L+2)] } \nonumber \\
&=& 1 - 2 { (2n+1) \sin {p \pi \over L+2} - \sin {(2n+1)p \pi \over
L+2} \over (2L+3) \sin {p \pi \over L+2}
- \sin {(2L+3)p \pi \over L+2} } \ .
\end{eqnarray}
We set $n=xL$ and work in the limit of infinite $L$.  Thereby we find
\begin{equation} \label{SimPro}
m_p(x) \equiv M_p (xL) =
1 - 2 x + {1 \over p \pi} \sin (2px \pi ) \ .
\end{equation}
In Fig. 3, $m_{1}(r)$ is shown together
with the numerically determined
ground-state profiles $m(r)$ for
$S=\frac{1}{2}$, $S=1$ and $S=\frac{3}{2}$
and we find a nice qualitative agreement (even for moderately large
values of $h$).

The above calculation can be generalized to larger $S$.
Consider a state in which the wall is between lattice sites.
In this state let all spins to the left of the wall have
$S_z=S$ and those to the right have $S_z=-S$. Note that
it is possible to change the value of $S_z$ for {\it either one}
(but not simultaneously both) of the spins adjacent to the wall
without changing the unperturbed ($h=0$) energy.  For spin $S=1$
since the nonzero matrix elements of the perturbation due to the
transverse field are all the same, the problem
is totally equivalent to a spin 1/2 chain of twice the length.
This observation explains the fact that our results for
$S=1/2$ and $S=1$ are indistinguishable.
When $S > 1$, one has to account for the fact that to move the
wall through one lattice constant involves matrix elements
which depend on the initial and final values of $S_z$.
This case will be considered later.

\subsection{Profile Function for the Oscillatory Phase}

The discussion has been so far
restricted to the ferromagnetic phase F.
The oscillatory phase is distinguished from
it by showing a non-monotonic
decrease of the correlation function [see Eq. (\ref{Corrs})]. How does
this behavior manifest itself on the
level of the magnetization profile
$m(r)$ ?

To answer this question, we display in Fig. 6A, for $S=\frac{1}{2}$,
$m(r)$ for $h=0.01$ and $\epsilon=0.5$ and for $L$ {\em even}.
Indeed, this profile
is distinct form the ones observed previously in the phase F. First,
finite-size effects are much larger than in the situations discussed
before. Second, $m(r)$ displays, at least for $L$ finite, a step-like
behavior and it looks as if the system was built out of hard objects
each occupying two lattice sites.
When $h$ is increased, these composites
soften until they completely melt
at the transition towards the F phase.

This picture is modified in interesting ways
for $L$ {\it odd}. In fact, the
calculation of $m(r)$ requires a little more care in this case.
For $L$ even, the system has a
single well-defined ground state separated
by a gap at least of order ${\cal O}(L^{-3})$ (see Sec. V)
from the excited states. That is not
so for $L$ odd. Rather, for $h=0$ but $\epsilon\neq 0$, one finds that
the ground state is twofold degenerate. That is a new degeneracy which
has nothing to do with the ferromagnetic
ordering of the system. Even when
$h\neq 0$, the two lowest states in
$\cal M$ remain much closer to each
other than with the other states, the latter one having gaps of order
${\cal O}(L^{-2})$. This new (near) degeneracy implies that one must
reconsider the calculation of the magnetization profile $m(r)$. Rather
than blindly using Eq.~(\ref{MagPro}), we take the {\em two}
ground states $|0\rangle$ and $|0'\rangle$ to be nearly
degenerate. Then we construct the matrix
\begin{equation} \label{OrdParMat}
{\bf m} (r) = S^{-1}
\left(
\begin{array}{cc}
{\langle0|S_z(n)|0\rangle} & {\langle0'|S_z(n)|0\rangle} \\
{\langle0|S_z(n)|0'\rangle}& {\langle0'|S_z(n)|0'\rangle}
\end{array}
\right) \;\; , \;\; r=n/(L+1)
\end{equation}
and find its two eigenvalues $m_{\pm}(r)$. Each of those represents
a magnetization profile and one of them is shown in Fig. 6B. The
other profile is obtained by reflection around $r=0.5$.
The asymmetry in Fig. 6B, observed for $L$ odd and finite, is
easily understood in terms of the composite objects introduced above.
Note that within the manifold ${\cal M}$, the $y-y$ term in the
Hamiltonian moves the wall through two lattice spaces.  Since
we are discussing $h=0$, this is the only kinetic energy of the
domain wall.  Thus, if we write ${\cal M}={\cal }M_{\rm e} +
{\cal M}_{\rm o}$, where ${\cal M}_{\rm e}$ (${\cal M}_{\rm o}$)
is the submanifold of state in which there are an even (odd) number
of up spins, then there are no matrix elements between these
two submanifolds.   Furthermore, these two submanifolds are
related to one another by spatial inversion (accompanied by
$S_z \rightarrow - S_z$).  Thus, for $L$ odd, the problem decomposes
into two identical eigenvalue problems (hence the twofold
degeneracy of the energies). The matrix to be diagonalized is exactly
the one considered in Sec. III.B, but now for $M=(L+1)/2$ sites and
with $h \rightarrow -\epsilon$. The two sets of eigenstates are thus
\begin{eqnarray}
\psi_p^{(-)} &=&
C \sum_{m=1}^{M} \sin\left(\frac{2\pi p m}{L+3}\right)
|2m-1> \nonumber \\
\psi_p^{(+)} &=&
C \sum_{m=1}^{M} \sin\left(\frac{2\pi p m}{L+3}\right)
|2m> \ ,
\end{eqnarray}
where $C$ is a normalization constant.
This eigenvector basis also renders
the matrix (\ref{OrdParMat}) diagonal. Writing $n=2k-1$ and $n=2k$,
respectively, we find for the profile for the $+$ set of eigenstates
(indicated by a subscript "$+$")
\begin{equation}
M_p^+(2k-1) = M_p^+(2k) =
1 - 2 \frac{(2k+1)\sin (2\pi p/(L+3))-
\sin( (2k+1)2\pi p/(L+3)) }{(L+2) \sin( 2\pi p/(L+3))-
\sin( (L+2)2\pi p/(L+3)) } \ .
\end{equation}
In particular, this reproduces the steps in $m_1^+(r)$ observed
numerically.  Now, taking the continuum limit, it is easy to see that
the width of the terraces decreases as $L\rightarrow\infty$
and that $\tilde m(x) \rightarrow m(x) +
{\cal O}(1/L)$. We therefore suggest that the phenomenon observed in
Fig. 7 is a novel type of finite-size effect.
As $L\rightarrow\infty$, the
step-like and asymmetric profiles found for $L$ finite will converge
towards a smooth and symmetric limit function.

At least for $h\rightarrow 0$, this limit function appears
to show a simple
relationship with the profiles previously
found in the F phase. Indeed,
our data suggest the following.
The profiles for $L$ large as found for
$h=0, \epsilon\neq 0$ are very close to those obtained for
$h=h_{\rm eff}(\epsilon ), \epsilon =0$.
For small $\epsilon$ our analytic work shows this to be
true with $h_{\rm eff} (\epsilon ) = \epsilon$. For
larger $\epsilon$ we have the phenomenological result,
\begin{equation}
h_{\rm eff}(\epsilon) \simeq \sqrt{\epsilon}
\end{equation}
Evidence for this is provided in Fig. 7, where some profiles found
for $h=0$ and $\epsilon\neq 0$ (data points) are collapsed with
profiles calculated with $h=h_{\rm eff}(\epsilon)$.

\section{PROFILES AT LARGE $S$}

In this section we obtain several analytic results in the
large $S$ limit.  We first analyze the
classical ($S \rightarrow \infty$) system.
Then we treat the quantum system with $S$ large, but
not infinite.

To treat the classical ($S=\infty$) system, we write
\begin{equation}
\label{CLASS}
S_x(n)/S = \cos \phi_n \ , \ \ \ \ \ \ \ S_z(n)/S = \sin \phi_n \ ,
\end{equation}
where $\phi$ is a continuous classical variable.
Then the energy of the classical system is
\begin{equation}
\label{ECLASS}
E = - {1 \over 2} H \sum_{n=1}^L \cos \phi_n -
{1 \over 2} J \sum_{n=1}^{L-1} \sin \phi_n
\sin \phi_{n+1} \ .
\end{equation}
For the classical calculation without a wall, we will adopt
periodic boundary conditions, so that $\phi_{L+1}=\phi_1$
and the last term in Eq. (\ref{ECLASS}) runs from $n=1$ to $n=L$.
In that case $\phi_i=\phi_{\rm eq}\equiv \arccos \tilde h$,
where, for convenience, we set
\begin{equation}
H/(2J)=h/2= h/h_c = \tilde h \ ,
\end{equation}
where $h_c=2$ is the critical value of $h$ above which
long-range Ising order disappears.  For the periodic chain
the energy per spin, $e_0$, is given by
\begin{equation}
e_0 = -{1 \over 2} J ( 1 + \tilde h^2) \ .
\end{equation}

In the classical calculations we will use two types of boundary
conditions to generate a wall.  The first type of boundary condition
is used for the continuum calculation.  Here
the number of sites is infinite and hence we set $\phi(x=-\infty)
=-\phi(\infty)=\phi_{\rm eq}$.  This continuum calculation is valid
when the angle $\phi$ changes slowly with position, as it does
for $\tilde h \approx 1$.  The second type of boundary condition
is that used in the discrete calculation.  Here it is convenient to
use antiperiodic boundary conditions in which spins at opposite ends
of the chain are coupled antiferromagnetically, with $K=J$ and we
require $\phi_{-i}=-\phi_i$.

In any case, the ground state energy is found by minimizing $E$
with respect to the $\phi$'s.
Apart from end effects this minimization
yields the following conditions (for $n=2, \ 3, \ 4, \dots N-1$)
which characterize the exact ground state,
\begin{equation}
\label{EXACT}
{1 \over 2} \Biggl( \sin \phi_{n+1} + \sin \phi_{n-1} \Biggr)
= (H/2J) \tan \phi_n \equiv \tilde h \tan \phi_n \ .
\end{equation}

\subsection{Classical Continuum Limit}

For small $\tilde h$ the angle $\phi$
will change abruptly at the domain
wall and we will treat this case
in the next subsection.  As $\tilde h$
increases (up to the critical value $\tilde h=1$), $\phi_n$ will
become a smoother function of $n$.
In that case, a continuum approximation
for $\phi_n$ makes sense.  To treat the continuum limit for
$\tilde h \approx 1$, we set
$\sin \phi_n = y(x) \equiv \sin \phi (x)$,
where $x=na$, whence Eq. (\ref{EXACT}) becomes
\begin{equation}
\label{EXACTY}
\tilde h y(x) = {1 \over 2}
\sqrt { 1 - y(x)^2 } [ y(x-a) + y(x+a) ] \ .
\end{equation}
The continuum limit of this equation is
\begin{equation}
\label{DIFEQ}
\tilde h y(x) = \sqrt { 1 - y(x)^2 } \left[
y(x) + {1 \over 2} a^2 {d^2y \over dx^2} \right] \ .
\end{equation}
The solution to this differential equation is reduced to quadratures:
\begin{equation}
\label{X}
{\sqrt 2 x \over a } = \phi(x) + \tilde h
{ 1 \over \sqrt{ 1 - \tilde h^2 } }
\ln \Biggl( { 1 - \tilde h \cos \phi(x)
+ \sqrt { 1 - \tilde h^2}
\sin \phi(x) \over \cos \phi (x) - \tilde h } \Biggr) \ .
\end{equation}

We now discuss the significance of this result.  First we
analyze the behavior for large $x$.  At large $|x|$, $\phi (x)$
approaches $\pm \phi_{\rm eq}$, where $\cos \phi_{\rm eq} =h$.
Thus for large $|x|$ we set
$\phi (x) =\pm [\phi_{\rm eq} - \delta(x)]$.
Keeping only the dominant terms in Eq. (\ref{X}), we obtain
\begin{equation}
{\sqrt 2 x \over a}- \phi_{\rm eq}(x) =
 - {\tilde h \over \sqrt {1 - \tilde h^2 } } \ln
\Biggl(  { \delta(x) \over 2 (1-\tilde h^2) } \Biggr)  \ .
\end{equation}
This gives
\begin{equation}
\delta (x) = \delta_0 e^{- x/\xi} \ ,
\end{equation}
where
\begin{equation}
\delta_0 = 2 (1 - \tilde h^2)
\exp \left( a \arccos \tilde h /( \sqrt 2 \xi )
\right)
\end{equation}
and $\xi$ is the correlation length, given by
\begin{equation}
{\xi \over a } = {\tilde h \over \sqrt 2 \sqrt {1 - \tilde h^2 } } \ .
\end{equation}
Thus, for $\tilde h \rightarrow 1$,
\begin{equation}
\xi /a \sim A_\xi (\tilde h_c - \tilde h)^{-\nu} \ ,
\end{equation}
with $\nu=1/2$, $\tilde h_c=1$, and $A_\xi = 1/2$.

Next let us see the behavior of the solution
near $x=0$.  To do that, differentiate the solution above with
respect to $x$ at $x=0$:
\begin{equation}
{\sqrt 2 \over a } = {d \phi \over dx } +
{ \tilde h \over 1 - \tilde h } {d \phi \over dx } \ ,
\end{equation}
where we used $\phi(0)=0$.  This gives
\begin{equation}
{d \phi \over dx } \biggr]_{x=0} = { \sqrt 2 \over a} (1-\tilde h) \ .
\end{equation}
To estimate the width, $W$, of the wall, we note that over a distance
$W$, the angle $\phi$ varies from approximately $-\phi_{\rm eq}$ to
approximately $+ \phi_{\rm eq}$.  This reasoning indicates that
\begin{equation}
\label{WALL}
W {d \phi \over dx } \biggr]_{x=0} = 2 \phi_{\rm eq} \ ,
\end{equation}
For $h \rightarrow 1$, we have
$\phi_{\rm eq} \approx \sqrt 2 \sqrt {1 - \tilde h}$, so that
\begin{equation}
W \approx 2a (1-\tilde h)^{-1/2} \approx 4 \xi \ .
\end{equation}
Thus for $h \rightarrow \tilde h_c$,
$W$ is of order $\xi$, as one might expect.

\subsection{Classical Wall At a Lattice Site}

In this and the next subsection we look at the discrete
equations, which are the appropriate ones for small $\tilde h$.
Since we are dealing with a classical system,
it is convenient to use antiperiodic boundary
conditions in which spins at opposite ends of the chain are
coupled antiferromagnetically, with $K=J$.

First we consider the case when the center of the wall is at a
lattice site.  To treat this case, we consider a chain with
an odd number of spins.  Number the sites
$-N, -N+1, \dots , -1, 0, 1, \dots , N-1, N$ and fix $\phi_0 = 0$.
Then $\phi_{-n}=-\phi_n$ and
the energy $E'$ for this chain of $2N+1$ sites is
\begin{equation}
E' = - J\sum_{n=1}^{N-1} \sin \phi_n \sin \phi_{n+1} -
{1 \over 2} J \sin^2 \phi_N - H \sum_{n=1}^N \cos \phi_n - {1 \over 2}
H \ .
\end{equation}
The $\phi_n$ are determined by Eq. (\ref{EXACT})
for $n=2, 3, \dots N-1$, by
\begin{equation}
{1 \over 2} \sin \phi_2 - \tilde h \tan \phi_1 =0 \ ,
\end{equation}
and by
\begin{equation}
\label{PHIN}
{1 \over 2} \sin \phi_{N-1} + {1 \over 2 } \sin \phi_N
- \tilde h \tan \phi_N = 0 \ .
\end{equation}
To obtain the solutions for the $\phi$'s at small $h$ we write
$\phi_n = \pi /2 - \epsilon_n$
for $\epsilon_n \ll 1$.
Then for $1<j<N$
\begin{equation}
\tilde h \cot \epsilon_1 = {1 \over 2} \cos \epsilon_2 \ ,\ \ \ \
\tilde h \cot \epsilon_j = {1 \over 2} [ \cos \epsilon_{j-1}
+ \cos \epsilon_{j+1} ] \ , \ \ \
\tilde h \cot \epsilon_N = {1 \over 2} [ \cos \epsilon_{N-1}
+ \cos \epsilon_N ] \ .
\end{equation}
To obtain the energy at order $\tilde h$
we only need to solve the equations
correct to order $\tilde h$, in which case
\begin{equation}
\epsilon_1=2\tilde h \ , \ \ \ \ \epsilon_2 =
\epsilon_3 \dots = \epsilon_N
= \tilde h \ .
\end{equation}
Then, to find the energy to order $\tilde h$,
we may set $\sin \phi_j=1$,
and $\cos \phi_j =(1 + \delta_{j,1}) \tilde h$, so that
\begin{equation}
-E'/J = (N-1) + \frac{1}{2} + \tilde h + {\rm O}(\tilde h^2) \ .
\end{equation}
The wall energy $E^{\prime}_w$ is given by the large $N$ limit of
\begin{equation}
E^{\prime}_w = E' - (2N+1)e_0 \ ,
\end{equation}
where $e_0$ is the energy per site of the uniform chain. Thus
\begin{equation}
E^{\prime}_w/J = -(N-\frac{1}{2}) -\tilde h + (N+\frac{1}{2})
+ {\rm O}(\tilde h^2) = 1-\tilde h + {\rm O}(\tilde h^2) \ .
\end{equation}

\subsection{Classical Wall Between Lattice Sites}

To treat the case when the center of the wall is midway between two
lattice sites, we consider a chain of $2N$ sites, numbered
$-N, -N+1 , \dots ,-2, -1, 1, 2, \dots N-1, N$.  We require
that $\phi_{-n}=-\phi_n$ and we use the same antiperiodic boundary
conditions as in the previous case.  Then the energy is
\begin{equation}
\label{EPRPR}
E'' = - J \sum_{n=1}^{N-1} \sin \phi_n \sin \phi_{n+1}
+ \frac{1}{2}J\sin^2 \phi_1 - \frac{1}{2}J \sin^2 \phi_N
- H \sum_{n=1}^N \cos \phi_n \ .
\end{equation}
As before, the wall energy, $E^{\prime \prime}_w$ is obtained via
\begin{equation}
E^{\prime \prime}_w = E'' - 2Ne_0 \ .
\end{equation}
The $\phi_n$ are determined by Eq. (\ref{EXACT}) for
$n=2, 3, \dots N-1$,
by Eq. (\ref{PHIN}), and by
\begin{equation}
\label{PHIREF}
{1 \over 2} (\sin \phi_2 - \sin \phi_1 )  = \tilde h \tan \phi_1 \ .
\end{equation}
To analyze these equations, set $\phi_n= \phi_{\rm eq} - \epsilon_n$.
Then, to first order in $\epsilon_n$ we have (for $1<n<N$)
\begin{equation}
\tilde h \tan \phi_{\rm eq} - {\tilde h \over \cos^2 \phi_{\rm eq} }
\epsilon_n = \sin \phi_{\rm eq} - {1 \over 2} \cos \phi_{\rm eq}
[ \epsilon_{n-1} + \epsilon_{n+1} ]
\end{equation}
or
\begin{equation}
2 \epsilon_n / \tilde h^2 = \epsilon_{n-1} + \epsilon_{n+1} \ .
\end{equation}
We expect an exponentially decaying solution for $\epsilon_n$ for
$n \gg 1$.  So set
\begin{equation}
\epsilon_n \sim e^{-na / \xi }
\end{equation}
so that
\begin{equation}
2/\tilde h^2 = e^{a/\xi} + e^{-a/\xi } \ ,
\end{equation}
from which we get
\begin{equation}
\label{DECAY}
e^{a / \xi} = {1 \over \tilde h^2} +
\sqrt { {1-\tilde h^4\over \tilde h^4}} \ .
\end{equation}
For small $\tilde h$, we get $\xi \approx a / (2 |\ln \tilde h |)$.
For $\tilde h \rightarrow 1$, we get $\xi= (a/2)(1-\tilde h)^{-1/2}$,
in agreement with the continuum result.  These results hold
irrespective of the position of the center of the wall.

We now analyze the solution near the wall for small $\tilde h$.  To do
that we first look at
Eq. (\ref{PHIREF}) for $\phi_1$ when
we set $\phi_j = \pi /2 - \epsilon_j$:
\begin{equation}
\tilde h \cot \epsilon_1 =
{1 \over 2} [ \cos \epsilon_2 - \cos \epsilon_1 ] \ .
\end{equation}
But $\epsilon_2 \ll \epsilon_1$, so
\begin{equation}
{ \tilde h \over \epsilon_1} = {1 \over 2}
\left[ 1 - \left( 1 - {1 \over 2}
\epsilon_1^2 \right) \right] = {1 \over 4 } \epsilon_1^2 \ .
\end{equation}
So
\begin{equation}
\label{EPS13}
\epsilon_1 = (4\tilde h)^{1/3} \ .
\end{equation}
{}From here on one should interpret $\tilde{h}$ to mean $|\tilde{h}|$.
Next, look at the equation for $\phi_2 = \pi /2 - \epsilon_2$:
\begin{equation}
\tilde h \cot \epsilon_2 = {1 \over 2}
[ \cos \epsilon_1 + \cos \epsilon_3 ]
\approx  1 \ .
\end{equation}
In this way we find that $\epsilon_n = \tilde h$, for $n>1$.

A more systematic approach shows that
we can write the solution for the $\epsilon$'s as
\begin{equation}
\label{SOLN}
\epsilon_1 = (4\tilde h)^{1/3}F_1[(4\tilde h)^{2/3}] \ , \ \ \ \ \
\epsilon_n =
\epsilon_{\rm eq} +
A_n(4\tilde h)^{2n-(7/3)}F_n[(4\tilde h)^{2/3}] \ ,
\ \ \ \ \  n > 1 \ ,
\end{equation}
where $A_n=2^{(6-5n)}$, $\epsilon_{\rm eq} = \sin^{-1} \tilde h$, and
the functions $F_n$ are analytic functions such that $F_n(0)=1$.

To calculate $E''$, note that, up to order $\tilde h^{4/3}$ in the
energy, we may write
\begin{eqnarray}
\sin \phi_1 &=& \cos \epsilon_1 =
1 - {1 \over 2} (4\tilde h)^{2/3} [F_1(\tilde h^{2/3})]^2 +
{1 \over 24} (4\tilde h)^{4/3} [ F_1(\tilde h^{2/3})]^4 \dots
\nonumber \\
\sin \phi_j & = & \cos \epsilon_j = 1
\ , \ \ \ \ \ \ j>1 \ ; \nonumber \\
\cos \phi_1 &=& \sin \epsilon_1 = (4\tilde h)^{1/3}
F_1 (\tilde h^{2/3}) + {\rm O}(\tilde h) \nonumber \\
\cos \phi_j & = & \sin \epsilon_j =
{\rm O}(\tilde h) \ , \ \ \ \ \ \  j>1 \ .
\end{eqnarray}
Putting these evaluations into Eq. (\ref{EPRPR}), we obtain
\begin{eqnarray}
- E''/J & = & 1 - \frac{1}{2}(4\tilde h)^{2/3} [F_1(\tilde h^{2/3})]^2
+ {1 \over 24} (4\tilde h)^{4/3} [ F_1(\tilde h^{2/3})]^4
+(N-2) \nonumber \\ && \ \
- \frac{1}{2}\biggl[ 1 - (4\tilde h)^{2/3} [F_1(\tilde h^{2/3})]^2
+ {1 \over 3} (4 \tilde h)^{4/3} F_1(\tilde h^{2/3})]^4 \biggr]
+ \frac{1}{2} + 2\tilde h [ (4\tilde h)^{1/3} F_1(\tilde h^{2/3})]
\nonumber \\
&=& N-1 + (3/8) (4\tilde h)^{4/3} + {\rm O}(\tilde h^2) \ .
\end{eqnarray}
where we used the definition that $F_1(0)=1$.  Then the wall energy is
\begin{equation}
E^{\prime \prime}_w/J = -(N-1) - (3/8) (4\tilde h)^{4/3} + N
= 1 - (3/8) (4\tilde h)^{4/3} \ .
\end{equation}
Note that this result is {\em not} identical to the case when
the center of the wall is at a lattice site.  The physics
of this result will be discussed in more detail in the next section.

\subsection{Quantum Domain Wall for Large $S$}

We now consider the quantum chain for large $S$.
For a chain of $L$ spins of magnitude $S$,
there are $2LS+1$ states which have the same $h=0$ ground state
energy and which therefore must be treated within degenerate
perturbation theory.  Numerical studies show that for $S>1$,
this manifold of states splits into bands
which are slightly different depending on whether $2S$ is even or odd.
For $S$ half integer
there are $2S$ bands, each having $L$ states except for the
middle band which has $L+1$ states.  Approximately, the centers of
these bands are at energies
\begin{equation}
\label{CENTER}
\frac{1}{2}(L-1)J + \frac{1}{2S} m H ,
\end{equation}
where $m$ assumes the values $0$, $\pm 3/2$,
$\pm 5/2$, $\dots$ $\pm S$.
For integer $S$ one has
$2S-1$ bands, each having $L$ states except for the middle band
which has $2L+1$ states.  In this case the centers of the bands
are given by Eq. (\ref{CENTER}), but $m$ assumes the values $0$,
$\pm 2$, $\pm 3$, $\dots$, $\pm S$.  As an example, in Fig. 8 we
show the density of states for $S=3$ within the manifold ${\cal M}$.
As Fig. 8 illustrates, the bands become wider and the gaps
between bands become smaller as one approaches zero energy
(about which the levels occur symmetrically).
A qualitative explanation of the occurrence of such band gaps
is as follows.  If the spin 1/2 case is likened to a hopping
model, then the spin $S$ case is analogous to a hopping model
in which there is a periodically variable hopping matrix element.
This problem is therefore analogous to that of an electron in a
weak periodic potential, where one knows\cite{ASHMER} that
even for arbitrarily weak potentials one has band gaps.

We will now demonstrate the existence of the band gaps and obtain
quantitative information on the band widths for $h \ll 1$ as follows.
We will give a construction for the outermost (and narrowest)
sub-band, for which $m=S$ in Eq. (\ref{CENTER}).  Consider the
submanifold of states, ${\cal M}_S$, which contains the states,
shown in Fig. 5, $|p\rangle$ (for $p=1, 2, \dots , L)$.
These states are defined so that all spins (if any) having $i<p$ have
$S_{iz}=S$, all spins (if any) having $i>p$ have $S_{iz}=-S$,
and $S_{px}=S$.  One may verify that the state $|n \rangle$ is in
the subspace ${\cal M}$ of states of minimum energy eigenstates,
$E_0$, of the $h=0$ Hamiltonian, i.e., that
$J/(2 S^2)\sum_{i=0}^{L+1}S_{iz}S_{i+1,z}| n \rangle
= E_0 | n \rangle$.
The states $|n \rangle$ are not orthogonal to one another,
but do have an overlap which is small for large $S$ and is given by
\begin{equation}
\langle p | p+1 \rangle =
\langle S_z=S|S_x=S \rangle^2 \equiv \tau \ ,
\end{equation}
for $0< p < L$.  Rose\cite{ROSE} shows that
\begin{equation}
\label{TUNNEL}
\langle S_z=S|S_x=S \rangle = D^{(S)}_{S,S}(0,\pi/2,0) = 2^{-S} \ ,
\end{equation}
where ${\bf D}$ is a rotation matrix.  Therefore, we use
states which are orthonormalized to leading order in $\tau$,
\begin{equation}
| \tilde n \rangle = |n \rangle - (\tau /2) |n+1 \rangle
- (\tau /2) |n-1 \rangle  \ ,
\end{equation}
where we interpret $|0 \rangle$ and $|L+1 \rangle$ to be zero.

We now work only to first order in $\tau$.  Then one
notes that unless $n-1 \leq k \leq n+1$,
$h S_{kx} |n \rangle$ has zero overlap with the subspace
${\cal M}$. To see this, note that
$\langle S_z = m |S_x | S_z =m\rangle=0$.
Therefore we write, correct to leading order in $\tau$,
\begin{eqnarray}
\label{HOFF}
\langle \tilde n & | & \tilde {\cal H } | \tilde n+1 \rangle
\nonumber \\  & = &
\langle n | \tilde {\cal H} | n+1 \rangle
- (\tau/2) \langle n | \tilde {\cal H} | n \rangle
- (\tau/2) \langle n+1 | \tilde {\cal H} | n+1 \rangle
\nonumber \\ & = & -\frac{H}{2S} \sum_{k=n-1}^{n+1}
\Biggl( \langle n | S_{kx}  | n+1 \rangle
- (\tau/2) \langle n | S_{kx} | n \rangle
- (\tau/2) \langle n+1 | S_{kx} | n+1 \rangle \Biggr) \nonumber \\
& = & -\frac{H}{2S} \Biggl(
\langle n | S_{nx} + S_{n+1,x}  | n+1 \rangle
- (\tau/2) \langle n | S_{nx} | n \rangle
- (\tau/2) \langle n+1 | S_{n+1,x} | n+1 \rangle \Biggr) \nonumber \\
& = & -\frac{H}{2S} \Biggl(
2S \langle n | n+1 \rangle
- (S\tau/2) \langle n | n \rangle
- (S\tau/2) \langle n+1 | n+1 \rangle \Biggr) \nonumber \\
& = & -\frac{1}{2} H \tau = -H 2^{-2S-1} \ ,
\end{eqnarray}
where $\tilde {\cal H} \equiv {\cal H} - E_0$.  Also
\begin{equation}
\label{HDIAG}
\langle \tilde n | \tilde {\cal H } | \tilde n \rangle =
\langle n | \tilde {\cal H} | n \rangle + {\rm O} (\tau^2)
= -\frac{H}{2S} \langle n | S_{nx} | n \rangle  = - \frac{H}{2} \ .
\end{equation}
Thus we expect this subspace of states with $S_x=S$ to form
a sub-band centered at energy
$E_0-H/2$
with a width determined
by the hopping matrix element\cite{BAND} $t(S) = -H2^{-2S-1}$.
Thus the width of this band should be
$4|t|=H 2^{1-2S}$.  Numerically, for $S=5$ we found
a band width of $0.002 H$ in excellent agreement with our
calculation.  We thus have an analytic description of the
outermost two sub-bands corresponding to $S_x = \pm S$.  We
did not consider an analysis of the remaining inner sub-bands.
Note that the results shown in Fig. 8 are not for asymptotically
large $S$.  We did check that when $S$ is large enough, the
outermost subband does become symmetric, as one would expect
for a one-dimensional density of states corresponding to
Eqs. (\ref{HOFF}) and (\ref{HDIAG}).

In a sense, one sees from this calculation that for the wall to move
one lattice spacing, it must tunnel through the phase space of
spin states indicated by the result of Eq. (\ref{TUNNEL}).
In this connection a nice analogy has been suggested by
Stinchcombe.\cite{RBS}  Consider a rotor in a strong $\cos (2 \theta)$
potential.  On knows from exact solutions of the associated
eigenvalue problem that one has a limit in which one has
harmonic oscillator levels in the bottom of the potential well,
all of which are doublets with a small tunnel splitting.\cite{ROTOR}
It is tempting to think that one can provide a field theory in
one space and time dimension in which the single potential
minimum of the rotor becomes a periodic potential in a lattice.
Then the bands would reflect the tunnel splitting.

\section{NATURE OF THE SPECTRUM}

\subsection{$L^{-2}$ Spectrum in the Ferromagnetic
Phase for Small $S$}

So far, we have considered the form of
the profiles $m(r)$ and have seen
that at least qualitatively, their form can be understood from simple
degenerate perturbation theory. The manifold
$\cal M$ considered consists
of those states which give the lowest energies
in the $h\rightarrow 0$ limit.
For the case of $S=\frac{1}{2}$, for $h=0$ these are the states
shown in Fig. 5, namely,
\begin{equation}
\left| \uparrow \uparrow \cdots \uparrow
\downarrow\right\rangle \;\; , \ldots , \;\;
\left| \uparrow \cdots \uparrow \downarrow
\cdots \downarrow \right\rangle \;\; ,
\ldots , \;\;
\left| \uparrow \downarrow \cdots \downarrow \downarrow \right\rangle
\end{equation}
and for $L$ sites, there are $L+1$ of them. For $S=1$ one starts at
$h=0$ from the states
\begin{eqnarray}
{}~& \left| \uparrow \uparrow \cdots \uparrow
\downarrow\right\rangle \;\; ,
\ldots , \;\;
\left| \uparrow \cdots \uparrow \downarrow \cdots
\downarrow \right\rangle \;\; ,
\ldots , \;\;
\left| \uparrow \downarrow \cdots \downarrow \downarrow
\right\rangle \nonumber \\
 & \left| \uparrow \uparrow \cdots \uparrow 0
\downarrow\right\rangle \;\; ,
\ldots , \;\;
\left| \uparrow \cdots \uparrow 0\downarrow \cdots
\downarrow \right\rangle \;\; ,
\ldots , \;\;
\left| \uparrow 0\downarrow \cdots \downarrow \downarrow \right\rangle
\end{eqnarray}
and one gets a subspace of $(L+1)+L =2L+1$ states.
All these states have
in common that they contain a single wall.
States which do not belong to this ``one-particle
subspace'' $\cal M$ have a large gap with the ground state. It is now
remarkable that this clear separation of the subspace $\cal M$ from
all other states remains
intact even for {\em finite} values of $h$. The subspace $\cal M$ has
a simple analogy for (anti)periodic boundary
conditions. In these cases,
$\cal H$ commutes with the (modified) translation operator
$\cal T$ (or $\tilde {\cal T}$) and is
thus decomposed into block diagonal form, the blocks being labeled
by the eigenvalues of $\cal T$ (or $\tilde {\cal T}$).
$\cal M$ corresponds to the
set of lowest eigenstates of the blocks of $\cal H$.

While the gap of the lowest excited state {\em outside} $\cal M$ with
the ground state is finite (e.g. $1-h$ for $S=\frac{1}{2}$
and when $\epsilon=0$),
the gaps inside $\cal M$ scale with
$L^{-2}$. If $g(i)=E_i - E_0$ denotes the $i^{\rm th}$ gap, we expect
the scaling $g(i) \sim L^{-\theta}$ with $\theta=2$,
as obtained in Eq. (\ref{PertTh}).
Finite-size estimates for $\theta$ are then
obtained from, with $N=L+2$
\begin{equation} \label{ThetaFS}
\theta_{L} = \frac{ \ln(g(i)_{N+1} / g(i)_{N})}{\ln(N/(N+1)) }
\end{equation}
The extrapolation towards the limit $N\rightarrow\infty$
was carried out
with the BST extrapolation algorithm\cite{Chri93}.
We illustrate the convergence of the finite-size data
with a few examples
for $S=1$ in table~\ref{Tab1}. The convergence towards
$\theta=2$ is even clearer for $S=\frac{1}{2}$.

We now consider the finite-size amplitudes
\begin{equation}
a(i) = N^2 g(i) \ .
\end{equation}
The motivation for this comes from systems being precisely at a
critical point. In that case, it is known that for one-dimensional
quantum chains finite-size amplitudes defined as $N g(i)$ are
related\cite{Card84} to the anomalous dimensions of the scaling
operators of the model and this leads to simple patterns of the
spectrum of the amplitudes
(see e.g. Ref. \onlinecite{Chri93} for more information).
Although we are not working here at a critical point,
we observe a simple structure of the
excited states within $\cal M$ for $L\rightarrow\infty$ out of the
critical region. Namely
\begin{equation} \label{AmpRat}
r(i-1) = a(i-1)/a(1) = \frac{1}{3} (i-1) (i+1)
\end{equation}
Evidence for this is presented in table~\ref{Tab2} for $S=\frac{1}{2}$
and table~\ref{Tab3} for $S=1$ which
give the BST extrapolated\cite{Chri93}
estimates for the $a(i)$. It is remarkable that the {\em same}
expression should hold for the gaps ratios of both the $S=\frac{1}{2}$
and the $S=1$ models, even though $h$ is not very small and/or
$\epsilon\neq 0$. For $h$ small and $\epsilon=0$, this result can be
recovered from the energies
(\ref{PertTh}) in lowest order of perturbation theory.

\subsection{$L^{-\theta}$ Spectrum in the Oscillatory
Phase for Small $S$}

Turning to the oscillatory phase, we note that the spectrum of the
states within the subspace $\cal M$ is different from that in the
F phase. Here there are important
differences between $L$ even and $L$ odd.

For simplicity, we restrict attention to spin $S=\frac{1}{2}$ and
first concentrate on the line $h=0$. The structure of the levels
we are going to find is shown in Fig. 9. For $L$ even, the spectrum
of the low-lying states corresponding to
$\cal M$ is grouped into doublets.
While the doublet splitting is of order ${\cal O}(L^{-3})$, the gaps
between doublets are of order ${\cal O}(L^{-2})$. For $L$ odd, on the
other hand, the energy levels are doubly degenerate, with gaps
between them of order  ${\cal O}(L^{-2})$.

To see this, we look for a scaling behavior
of the energy gaps $g(i)=E_i - E_0 \sim L^{-\theta}$ and define
finite-size estimates for the exponent $\theta$ using (\ref{ThetaFS}).
In table~\ref{Tab5}, we give our results for $\epsilon=0.02$. For $L$
even, estimates of $\theta$ from the lowest six gaps ($i=1,\ldots,6$)
are given.  (For some of the higher gaps, the finite-size data for $L$
small are not yet in the asymptotic regime and are thus discarded.)
We observe that the lowest gap scales with an exponent $\theta=3$,
while all the higher gaps scale with $\theta=2$. For $L$ odd, on the
other hand, each energy level is doubly
degenerate even for finite $L$,
i.e. $g(2j)=g(2j-1)$, $j=1,2,\ldots$.
Estimates of $\theta$ for $i=2,4,6$ are given in
table~\ref{Tab5} and we find $\theta=2$
throughout. Similar results are also found for
$\epsilon=0.1$ and $0.5$.

Next, we study the finite-size amplitudes $a(i)=N^2 (E_i - E_0)$, with
$N=L+2$. Estimates for the lowest six amplitudes extrapolated to
$L\rightarrow\infty$ are shown
in table~\ref{Tab6}, obtained for $L$ even or
odd, respectively. We find
that for $L$ even, the estimates for pairs of
amplitudes are quite close
to each other and are consistent with
being equal. Furthermore, their numerical
values are near to the ones found for $L$ odd. Taken together, the
present data suggest the simple picture that in
the $L\rightarrow\infty$ limit,
the amplitudes $a(i)$ should become doubly
degenerate and independent of
the evenness or oddness of $L$. Finally,
when looking for the amplitude
ratios $a(i)/a(2)$, we find the following pattern
\begin{equation}
0, 1, 1, \frac{8}{3}, \frac{8}{3}, 5, \ldots
\end{equation}
which is consistent with the very same structure (\ref{AmpRat})
of the low-lying amplitudes
found on the $\epsilon=0$ line.
This finding is certainly consistent with
the relationship between the order parameter profiles observed earlier
(see Fig. 7) between systems on the $h=0$ and $\epsilon=0$ lines.

These findings can be reproduced from the perturbative arguments
of section III.B when $\epsilon$ is small.
For $L$ odd, the eigenvalue problem
then reduces to one treated there, but with $(L+1)/2$ sites and
$h\rightarrow -\epsilon$ and each state being
twofold degenerate. For $L$ even,
we get two distinct problems, one
with $L/2$ sites and one with $L/2+1$
sites. Using the expression (\ref{PertTh}) for
the energies, we find pairs
of levels with a splitting of order ${\cal O}(L^{-3})$ between them.

Qualitatively, the same structure persists throughout the O phase.
We illustrate this in Fig.~10, where we show for $\epsilon=0.25$ and
$L=10$ sites the dependence of the first few eigenvalues on $h$.
Because $L$ is not very large, the statements we shall make about
the spectrum apply most strongly to the lowest levels and less well
to the higher levels. We see
that inside the O phase, the lowest energies occur in pairs which
oscillate around another. In the ground state, there are for $L$ sites
$(L+1)/2$ level crossings for $L$ odd and $L/2$ level crossings for
$L$ even, respectively. These level crossing do not continue into
the F phase. Finally, we point out
that at the F/D transition, the energies
combine into a $\left(\frac{1}{2}\right)$
representation of the $c=\frac{1}{2}$
Virasoro algebra, as predicted from
conformal invariance\cite{Chri93,Card86}.
In agreement with that prediction, we observe in Fig. 10 close
to the critical point $h_c=1+\epsilon$ an (approximately)
equidistant level spacing with the
degeneracies $1,1,1,1,2,\ldots$.
(Actually, because of finite size effects, this structure is
realized at $h=1.15$ rather than at $h=h_c=1.25$.)

This kind of level crossings in the ground state of finite-lattice
quantum system has been first investigated
for {\em periodic} boundary
conditions\cite{Hoeg85}.
There is was noted that the level crossings always
occurred between the $Z_2$
even and odd sectors of the model. Here we find
that the level crossing persist even if
the $Z_2$ symmetry is broken by
our chosen ``up-down'' boundary conditions (\ref{BC}).
Applying finite-size
scaling\cite{Hoeg85} to the location $h_k$ of the level crossings,
we expect (but did not test) a scaling
$h_k (L) - h^{*} \sim L^{-1/\nu}$.  For periodic boundary
conditions it is known\cite{Hoeg85} that $\nu=1/2$.
($h^*= \sqrt{4\epsilon}$ gives the O/F transition line).

\subsection{Intrinsic Wall Width}

It is obvious that the fact that the width of the wall for
the quantum model is of order the system size is a result
of a quantum superposition of states each of which have a
narrow wall.  It is tempting, therefore, to introduce a
measure of the ``intrinsic" width of the wall, which is
the minimum width obtainable within the subspace of
states under consideration.  We emphasize that this
concept depends on the subspace of states begin considered.
As we vary $\epsilon$, the strength of the $y$-$y$ coupling,
it seems plausible that the character of the low lying
``particle" states may vary.  Thus, it would be of
interest to introduce a measure of the intrinsic width
applicable to this case.  For this purpose consider the
quantity $Q$ defined as
\begin{equation}
\label{MAX}
Q = (2S)^{-1} {\rm Max}_{n \in \cal M}
\langle n | S_z(i) - S_z (i+1) |
n \rangle \ ,
\end{equation}
where the maximum is to be taken over all possible quantum states
$|n \rangle$ in the subspace ${\cal M}$.  (As long as the site
$i$ is not near an end, $Q$ will depend only weakly on $i$.)
If the states consist of sharp walls, then $Q$ will be unity.
So we use this quantity to define the intrinsic width, $W_i$,
via $W_i \sim 1/Q$.  This type of definition is somewhat
similar to the inverse participation ratio introduced to
characterize localized and extended states of an electron
in a random potential.\cite{RANDOM}  Here we take the
subspace ${\cal M}$ to be the ``particle" subspace of states.
This definition is appropriate as long as the subspace of
particle states can be unambiguously separated from the
continuum.  This requirement suggests that $h \ll J$.
But, we can also use this idea when $\epsilon$ is nonzero.

We should mention that the maximization required in Eq. (\ref{MAX})
is easy to carry out.  One simply forms the matrix, ${\bf P}$,
(in any representation) where
\begin{equation}
{\bf P}_{n,m} = (2S)^{-1}
\langle n | S_z(i) - S_z (i+1) | m \rangle \ ,
\end{equation}
and $Q$ is the largest eigenvalue of ${\bf P}$.
When $Q$ is less than unity, one also obtains information
on the shape of the wall via the eigenvalue spectrum of
${\bf P}$.  For instance, if the wall has width 3, and
if the three largest eigenvalues of ${\bf P}$ are
degenerate, then one would conclude that the intrinsic wall
profile is linear.

One context in which this concept provides some information is
when we consider the large $S$ limit of the transverse Ising
model.  In that case we have already seen that the low lying
particle spectrum splits into bands corresponding roughly
to values of $S_x$.  When we apply Eq. (\ref{MAX}) taking
the subspace ${\cal M}$ to be the lowest band of $L$ states with
$S_x=S$, we expect to find $Q=1/2$.  We in fact verified this
expectation in our numerical calculations.  They showed that
one obtains two nearly degenerate maximal eigenvalues of
magnitude essentially equal to $1/2$.  There are two eigenvalues
because there are two wavefunctions which can maximize $Q$
since the minimum intrinsic wall width is two lattice spacings.
We see then, that the prescription of Eq. (\ref{MAX}) coincides
with the perturbative calculations for small $h$.

\subsection{Barrier to Domain Wall Motion}

First we summarize the result for the classical
($S \rightarrow \infty$)
system.  Note that $E_w^{\prime}
< E_w^{\prime \prime}$, for small $H$.
So to move the
wall through one or more lattice constants requires an energy
\begin{equation}
\label{BARRIER}
\Delta = E^{\prime \prime}_w - E^\prime_w = J \biggl[ \tilde h
- (3/8) (4 \tilde h )^{4/3} \biggr] \ .
\end{equation}
We have numerically studied the energies of the classical
domain wall when its center is either at a lattice site or midway
between lattice sites.  We find that the barrier energy $\Delta$
does not change sign as $H$ is increased as might be suggested by
Eq. (\ref{BARRIER}), but rather $\Delta$ decreases monotonically
to zero as $H$ is increased towards its critical value of unity.

We carried out similar calculations for quantum systems
with $S=\frac{1}{2}$ and $S=1$. We found that the difference in
energy when the center of the wall moved from a lattice
site to a point midway between lattice sites was
too small to be accessible to double precision arithmetic.
Since, for a chain of length $L<20$, we can not imagine a
parameter on such a scale, we assert that the quantum barrier
energy is zero.
Practically, we performed this test as follows. For sufficiently
large lattice sizes $L$, we expect for the ground state energies
$E_0(2N)$ (i.e. wall between two sites) and $E_0(2N+1)$ (i.e. wall
at a site)
\begin{equation}
\left\{
\begin{array}{c} {E_0(2N) = A\cdot 2N + B + \ldots } \\
{E_0(2N+1) = A\cdot(2N+1) + B' + \ldots } \end{array}
\right. \ ,
\end{equation}
where $A$ is the ground state energy per spin
and independent of the boundary conditions.
For $S=\frac{1}{2}$, $A=-\frac{1}{\pi}(h+1)E(\sqrt{4h}/(h+1))$
where $E$ is a complete elliptic integral\cite{Pfeu70}.
The desired wall energy difference should be proportional to
$B-B'$. This in turn can be estimated from the quantities
\begin{eqnarray}
p_N &=& E_0(2N+1)+E_0(2N-1) - 2 E_0(2N) \simeq 2 (B'-B) + \ldots
\nonumber \\
q_N &=& E_0(2N+2)+E_0(2N) - 2 E_0(2N+1) \simeq 2 (B-B') + \ldots
\end{eqnarray}
and we note that $p_N$ and $q_N$ should have opposite signs.
We give in table~\ref{Tab4} estimates
for both $p_N$ and $q_N$ for both
$S=\frac{1}{2}$ and $S=1$ for $h=0.1$.
We chose this value of $h$ to be small enough that is was for from
criticality but large enough so that $|B-B'|$ should be easily
observable if nonzero.
Taking the entries at face value, the wall energies must be less than
${\cal O}(10^{-5})$. Moreover, since $p$ and $q$ should have opposite
signs, the values given in table~\ref{Tab4} appear to be entirely due
to the leading finite-size corrections and we find that $p_N, q_N$
decrease rapidly with increasing $N$. So we get an upper bound
\begin{equation}
\mbox{\rm barrier energy} \sim |B-B'| \leq {\cal O}(10^{-8})
\end{equation}
for both $S=\frac{1}{2}$ and $S=1$.   This result is not
surprising.  In the limit when the width of the domain wall
diverges (as it does for $L \rightarrow \infty$), the barrier
energy should vanish.  Whether it vanishes algebraically
or exponentially would require further
numerical or analytic analysis.

\subsection{Crossover From Quantum to Classical Behavior}

Here we discuss the various regimes which exist at
at large $S$.  It is clear that bands of energy levels exist
on various scales.  It is useful to discuss the consequences
of these energy scales in terms of scaling functions.  For
notational simplicity, in this subsection $F(x)$ will denote
a scaling function of $x$ which, in general, will be different
in different appearances.

We first discuss the thermodynamic properties of the wall.
For this purpose we will consider the entropy associated
with the wall, $S_w$, defined as the entropy with ``up-down''
boundary conditions minus that with ``up-up'' boundary
conditions.  We emphasize that the scaling of the bulk
thermodynamics, i.e., the extensive part of the thermodynamic
potentials is independent of the boundary conditions and only
depends on the bulk thermodynamic variables.\cite{BULK}
At temperatures small compared to $J/S$ the
entropy with ``up-up'' boundary conditions will be zero,
because the ground state is nondegenerate.  The entropy
with ``up-down'' boundary conditions will be that due to
the band of states with just one domain wall.  First there
is an extreme low--temperature limit in which the entropy
is vanishingly small:
\begin{equation}
S_w \approx 0 \ ; \ \ \ \ \ \ T \ll H2^{1-2S}L^{-2} \ ,
\end{equation}
where we always set $k_B$, the Boltzmann constant equal to unity.
In this limit only the lowest single state of the tunneling band
found in Sec. IVB comes into play.  At higher temperatures the
tunneling band is activated, so that
\begin{equation}
S_w = \ln L + F\Bigl( {T \over H2^{1-2S} } \Bigr) \ ; \ \ \ \ \ \
H2^{1-2S}L^{-2} \ll T \ll H/S \ .
\end{equation}
Here $\ln L + F(x)$ is the entropy of a one dimensional tight binding
band of $L$ sites.  By noting the form of the density of states one
can see that
for small $x$, $F(x) \sim (1/2) \ln x + {\rm Const}$ and as
$x \rightarrow \infty$, $F(x) \rightarrow 0$.  The upper limit of
this regime is defined so that essentially only the lowest subband
of Sec. IVB is activated.  Next,
there is a regime when the temperature
is high enough that all states of
these subbands are equally populated,
but only one domain wall exists.
In this discussion it is assumed that
$H \ll J/S$, so that the band of particle
states is well separated from
the states with more than one wall.  In this regime one has
\begin{equation}
S_w = \ln (2SL) + F\Bigl( {T \over H } \Bigr) \ , \ \ \ \ \
H/S \ll T \ll J/S \ .
\end{equation}
In this case, $F(x) \rightarrow 0$ for $x \rightarrow \infty$
and $F(x) \sim \ln (T/2H) + 1$ for $x \rightarrow 0$.  Finally,
we have the classical regime when $T \gg J/S$.  In that regime
any effects due to the fact that $S_x(i)$ and $S_z(i)$ do
not commute with one another becomes unimportant.  In that
regime then, the partition function for the quantum model of
Eq. (\ref{HAMZZ}) is the same
[up to corrections of relative order $J/(ST)$]
as that of the classical model in which the
operators are interpreted as classical variables according to
Eq. (\ref{CLASS}).

The effect of a random potential is somewhat different.
In the present context a random potential would be created by
a field which independently for each site assumes random values
from a distribution of width $V_R$.  It is known\cite{Thouless}
that in the presence of a random potential all states are
localized.  That means that in the presence of a random field
the eigenfunctions are localized and therefore the width of the
wall in the ground state remains finite in
the limit $L \rightarrow \infty$.
Nevertheless, the width of the wall, $W$, can be expressed in
terms of scaling functions of the same variables as describe
the thermodynamics except that here $V_R$ replaces $T$.  Thus
\begin{equation}
W \sim L \ ;  \ \ \ \ \ \ V_R \ll H2^{1-2S}L^{-2} \ ,
\end{equation}
and
\begin{equation}
W/a = F\Bigl( {V_R \over H2^{1-2S} } \Bigr) \ ; \ \ \ \ \ \
H2^{1-2S}L^{-2} \ll V_R \ll H/S \ .
\end{equation}
Since we know that $W$ can not be less than the intrinsic
width, we see that $F(\infty)=2$.  Similarly, we have
\begin{equation}
W/a = F\Bigl( {V_R \over H } \Bigr) \ ;
\ \ \ \ \ \ H/S \ll V_R \ll J/S \ .
\end{equation}
In this case $F(0)=2$ and $F(\infty)=1$, the intrinsic wall width
in this regime.  For any larger $V_R$, $W/a$ remains equal to unity.

\section{SUMMARY}

We may summarize our conclusions for the Ising model
in a transverse field as follows.

\begin{itemize}
\begin{enumerate}
\item
Although an exact analysis of domain walls in the transverse Ising
model can in principle be accomplished using the Jordan-Wigner
transformation, it is much simpler to consider the nature of
degenerate perturbation theory in the presence of ``up-down''
boundary conditions.  Within degenerate perturbation theory for
$S=1/2$ and $S=1$ the Hamiltonian governing the position of the
center of the domain wall is isomorphic with a hopping model with
a hopping matrix element $t \sim H$.  We verified this in detail
with analysis of the scaling of the energy levels with chain length,
$L$.

\item
For spins $\frac{1}{2}$, $1$ and $\frac{3}{2}$ the ground state
profiles when spins at opposite ends of a
chain of length $L$ are fixed to be antiparallel show relatively
little dependence on the transverse field $h$ (as long as $h$ is
less than the critical value, $h_c$, above which ordering in $S_z$ is
destroyed).  For small $h$ the profiles at large but finite $S$ are
not very different from those at small $S$.  In all cases, the
profiles are quite well obtained from the hopping model.

\item
The profiles in the oscillatory phase show much larger finite size
corrections than those in the ferromagnetic phase.  In the
limit $L \rightarrow \infty$ the profiles in the oscillatory
phase may still show some oscillatory behavior, unlike in the
ferromagnetic phase.  Also, our results for the profiles show
an interesting finite-size effect in which spatial parity appears
to be broken.

\item
In the spectrum of the oscillatory phase at $h=0$ one finds an
unusual pattern for the gaps between the ground state and the
first few excited states.  While for $L$ even, the energies form
doublets, each with a splitting which scales with $L$ as $L^{-3}$,
and with gaps between the doublets which scale as $L^{-2}$. For $L$
odd, the energies are doubly degenerate and the gaps scale with
$L$ as $L^{-2}$.
This behavior was explained in terms of dimer excitations
which occur in the limit $h \rightarrow 0$ when $\epsilon \not= 0$.

\item
For large $S$ the quantum transverse Ising model the Hamiltonian
describing the position of the center of the domain wall gives rise
to band splittings analogous to those found for an electron in a
periodic potential\cite{ASHMER}.  The lowest sub-band
is described by a hopping model in which the center of the domain
wall tunnels from one site to a neighboring sites.  The
associated tunneling matrix element
is of order $hS \exp (- \alpha S)$,
where $\alpha = 2 \ln 2$.  It would be interesting to recover this
result within a field theory for one space and one time dimension.

\item
For the classical ($S=\infty$) model the domain wall is extremely
narrow for small $h$.  We have determined the domain wall energy
for the case when i) the center of the domain wall is at a lattice
site and ii) the center of the domain wall is midway between two
lattice sites.
The difference between these energies is the pinning energy
which must be exceeded to move the domain wall through the lattice.
For the quantum model we find this pinning energy
vanishes as $L \rightarrow \infty$ due to the fact that the width
of the domain wall diverges in this limit.

\item
In attempting to relate quantum problems to classical problems
it is necessary to define an intrinsic width, $W_i$, of a quantum
domain wall.  We have defined $W_i$ as the minimum possible
width attainable using wave functions from a manifold ${\cal M}$.
At temperatures large compared to the band but small compared
to the activation energy for creation of an additional wall,
the manifold ${\cal M}$ is the hopping band.  In that case, the
intrinsic width of the quantum system is essentially equal to
the width of the analogous classical domain wall.

\end{enumerate}
\end{itemize}

\vspace{0.5in}
\noindent
{\bf Acknowledgements}
The authors appreciate helpful advice from D.B. Abraham,
R. B. Stinchcombe, T. C. Lubensky, I. Affleck and
J.L. Cardy.
ABH thanks the Department of Theoretical Physics of Oxford
University for its kind hospitality.
ABH is supported by an EPSRC Visiting Fellowship and MH is
supported by a grant from the
EC programme ``Human Capital and Mobility".
MC acknowledges support by the Polish
Agency KBN (grant number 2 P302 127 07).

\newpage

\begin{table}
\caption[Exponent $\theta$]{Finite-size estimates of the
exponent $\theta$ for the
spin--1 model for two values of $h$.
The line labeled $\infty$ gives the $L\rightarrow\infty$
estimates and the numbers in brackets give the estimated
uncertainty in the last given digit.
\label{Tab1} }
\begin{tabular}{|c|ccc|ccc|} \hline
$\epsilon$ & \multicolumn{3}{c|}{$h=0.70711$} &
\multicolumn{3}{c|}{$h=0.141421$} \\ \hline
  5 & 2.1178266444 & 1.8259262012 & 1.6349745596
& 1.9232489201 & 1.5981924835 & 1.5459794089 \\
  6 & 2.1844497668 & 1.9905134772 & 1.7373695512
& 2.0506897828 & 1.8186085947 & 1.5210107108 \\
  7 & 2.2059303628 & 2.0578126044 & 1.9022922190
& 2.1146816632 & 1.9229956697 & 1.7423552051 \\
  8 & 2.2066110482 & 2.0951905179 & 1.9690586873
& 2.1430516979 & 1.9914094495 & 1.8343490446 \\
  9 & 2.1989322892 & 2.1141764054 & 2.0126854700
& 2.1534693122 & 2.0339467283 & 1.9014943987 \\
 10 & 2.1883357907 & 2.1226007841 & 2.0407754063
& 2.1549673204 & 2.0598773753 & 1.9493182117 \\
 11 & 2.1771411322 & 2.1251001574 & 2.0585469560
& 2.1520791521 & 2.0754047064 & 1.9830416672 \\
 12 & 2.1663136392 & 2.1243123257 & 2.0695528139
& 2.1471130353 & 2.0844025361 & 2.0068020462 \\
 13 & 2.1562240067 & 2.1217337201 & 2.0761282137
& 2.1412643703 & 2.0892668405 & 2.0235701176 \\
$\infty$& 2.0000(1)& 2.000(2)     & 2.01(2)
& 2.000(2)     & 2.00(2)      & 2.06(8) \\ \hline
\end{tabular}
\end{table}

\begin{table}
\caption[FSS-Amps for $S=1/2$]{Extrapolated estimates
for the finite-size scaling amplitudes $a(i)$ for $S=\frac{1}{2}$ in
the ferromagnetic phase F. The last column gives the amplitude
ratios $r(i)=a(i)/a(1)$ from Eq. (\ref{AmpRat}).  \label{Tab2} }
\begin{tabular}{|c|llllll|l|} \hline
         & $h=0.1$ & $h=0.1$ & $h=0.2$ & $h=0.2$
& $h=0.5$ & $h=1$ & \\
 Gap Nr. & $\epsilon=0$ & $\epsilon=0.0025$ &
           $\epsilon=0$ & $\epsilon=0.01$ &
           $\epsilon=0$ & $\epsilon=0.2$ & $r(i)$ \\ \hline
1 &  1.64492(2) &  1.480441(3) &  3.70110(1) &  2.960882(1) &
  14.803(3) & 29.272(3) & 1 \\
2 &  4.38641(2) &  3.94784(1)  &  9.8695(4)  &  7.895688(2) &
  39.406(2) & 79.998(40) & $\frac{8}{3}$ \\
3 &  8.2247(1)  &  7.40221(2)  & 18.5054(3)  & 14.804388(3) &
  73.865(8) & 146.31(10) & 5 \\
4 & 13.1595(8)  & 11.8435(1)   & 29.604(8)   & 23.68675(8)  &
 118.2(2)   & 233.9(5)   & 8 \\
5 & 19.190(2)   & 17.2719(2)   & 43.168(4)   & 34.542(1)    &
 172.5(2)   &            & $\frac{35}{3}$ \\
6 & 26.312(5)   & 23.687(2)    & 59.13(2)    & 47.371(3)    &
 236.5(3)   &            & 16 \\ \hline
\end{tabular}
\end{table}

\begin{table}
\caption[Amplitudes, spin 1]{Extrapolated estimates for the
finite-size amplitudes $a(i)$ for spin $S=1$. \label{Tab3} }
\begin{tabular}{|c|cccc|} \hline
 & \multicolumn{4}{c|}{$h$} \\ \hline
$i$ & 0.007071  & 0.035355   & 0.070711   & 0.141421   \\ \hline
1   & 0.0377469 & 0.20378(8) & 0.44676(3) & 1.0697(10) \\
2   & 0.10066   & 0.5434(6)  & 1.19141(15)& 2.866(10)  \\
3   & 0.1888(2) & 1.019(5)   & 2.234(5)   & 5.35(2)    \\
4   & 0.302(1)  & 1.67(5)    & 3.58(8)    & 8.55(10)   \\
5   & 0.444(4)  &            &            &            \\
6   & 0.605(8)  &            &            &            \\ \hline
\end{tabular}
\end{table}

\begin{table}
\caption[Wall energies]{Estimates for $10^5 p_N$ and
$10^5 q_N$, giving
upper bounds for the wall energy differences
$2|B-B'|$ at $h=0.1$ \label{Tab4} }
\begin{tabular}{|l|cc|cc|}\hline
 & \multicolumn{2}{c|}{Spin $\frac{1}{2}$} &
\multicolumn{2}{c|}{Spin 1}
                                   \\ \hline
$N$  &  $p_N$ &  $q_{N}$  & $p_N$  &  $q_N$  \\ \hline
2 & 224.4 & 124.0 & 102.1 &  48.9 \\
3 &  73.9 &  46.7 &  26.2 &  15.3 \\
4 &  30.9 &  21.3 &   9.5 &   6.2 \\
5 &  15.1 &  11.0 &   4.2 &       \\
6 &   8.2 &   6.3 &       &       \\
7 &   4.8 &   3.8 &       &       \\
8 &   3.0 &   2.5 &       &       \\ \hline
\end{tabular}
\end{table}

\begin{table}
\caption[Theta in O phase]{Finite-size data for the exponent $\theta$
obtained for $S=\frac{1}{2}$ with $h=0$ and $\epsilon=0.02$ and the
extrapolated limit for $L\rightarrow\infty$.
Estimates obtained for $L$ even
and odd, respectively, are shown separately. \label{Tab5} }
\begin{tabular}{|c|cccccc|c|ccc|}\hline
$L$ & \multicolumn{6}{c|}{ ~ } & $L$ &
\multicolumn{3}{c|}{ ~ } \\ \hline
  6 & 2.45042 & 1.19367 & 1.37847 & 0.80651 &         &         &
  5 & 1.01958 &         &          \\
  8 & 2.58995 & 1.38695 & 1.65858 & 1.13219 & 1.12909 & 0.74787 &
  7 & 1.36641 & 0.92322 &          \\
 10 & 2.67304 & 1.50917 & 1.78977 & 1.32647 & 1.44186 & 1.05893 &
  9 & 1.54228 & 1.25898 & 0.83163  \\
 12 & 2.72829 & 1.59281 & 1.86194 & 1.45465 & 1.61033 & 1.25543 &
 11 & 1.64710 & 1.44770 & 1.15571  \\
 14 & 2.76770 & 1.65329 & 1.90568 & 1.54487 & 1.71252 & 1.38993 &
 13 & 1.71580 & 1.56697 & 1.35237 \\
 16 & 2.79721 & 1.69886 & 1.93399 & 1.61140 & 1.77947 & 1.48709 &
 15 & 1.76383 & 1.64816 & 1.48286 \\
    &         &         &         &         &         &         &
 17 & 1.79902 & 1.70640 & 1.57477 \\ \hline
$\infty$ & 2.995(5) &2.00(2)&2.01(2)&1.98(3)& 2.01(1) & 1.99(3) &
$\infty$ & 1.999(2) &1.99(1)&2.00(1) \\ \hline
\end{tabular}
\end{table}

\begin{table}
\caption[Amplitudes in O phase]{Extrapolated finite-size amplitudes
$a(i)$, $i=1,2,\ldots,6$ for $S=\frac{1}{2}$ with $h=0$ and
for several values of $\epsilon$. For a given $\epsilon$,
the first (second) line
corresponds to the estimates found for $L$ even (odd).
\label{Tab6} }
\begin{tabular}{|c|cccccc|}\hline
$\epsilon$ & 1 & 2          & 3       & 4         & 5
& 6        \\ \hline
0.02   & 0 & 1.2033(20) & 1.22(4) & 3.191(20) & 3.243(15)
& 5.97(10) \\
       & 0 & 1.203(3)   & 1.203(3)& 3.191(15) & 3.191(15)
& 5.975(20)\\ \hline
0.10   & 0 & 6.553(15)  & 6.58(3) & 17.26(15) & 17.59(10)
& 32.3(3)  \\
       & 0 & 6.555(20) & 6.555(20)& 17.32(8)  & 17.32(8)
&32.48(10) \\ \hline
0.50   & 0 & 57.2(9)    & 58.8(6) & 153(4)    & 158(3)
& 311(9)   \\
       & 0 & 58.3(3)    & 58.3(3) & 156(2)    & 156(2)
& 325(10)  \\ \hline
\end{tabular}
\end{table}
\newpage

\begin{center}
{\large\bf FIGURE CAPTIONS}
\end{center}
\vspace{0.5cm}

\newcounter{fig}
\begin{list}{Fig. \thefig}{\usecounter{fig}\labelwidth32pt}

\item
Phase diagram of the ``yz" model of Eq. (\ref{YZEQ}) for
spin $S=\frac{1}{2}$. The disordered (D),
ferromagnetic (F) and oscillatory (O) phases are indicated.

\item
Local magnetization profile $m(r)$ for the spin-$\frac{1}{2}$ model
with $h=0.1$ and $\epsilon=0$ for system sizes $L=2 \ldots 16$.

\item
Comparison of the magnetization profiles $m(r)$ for the
spin-$\frac{1}{2}$ (boxes), the spin-1 (full circles) and the
spin-$\frac{3}{2}$ (open circles) models. The full
curve gives the profile of Eq.~(\ref{SimPro})
which is correct to first order
in $h$. The data are for $\epsilon=0$ and
A) $h=0.1$ and B) $h=0.5$.

\item
Magnetization profile $m(r)$ for three values of the transverse
field $h$ and for $\epsilon=0$ and $S=\frac{1}{2}$.

\item
States in the ground manifold ${\cal M}$.  Top: the states
$|n \rangle$ introduced for spin 1/2.  Bottom: the states
$|n \rangle$ introduced for the case $S \gg 1$.  In this case
the $n$th spin has $S_x=S$.

\item
Magnetization profile of the oscillatory phase of
the spin-$\frac{1}{2}$
model with $\epsilon=0.5$ and $h=0.01$,
for A) $L$ even and B) $L$ odd.

\item
Comparison of the magnetization profiles of the
spin-$\frac{1}{2}$ model
with $h\rightarrow 0$ and the values of
$\epsilon$ indicated (data points) and
with $h=h_{\rm eff}(\epsilon)=\sqrt{\epsilon}$ and $\epsilon=0$
(curves), for $L=15$.

\item
Density of states (per site) $D(E)$ as a function of energy $E$
(relative to the center of the manifold)
within the manifold ${\cal M}$ for $S=3$.  The middle subband
contains $2L+1$ states.  The other subbands contain $L$ states.
The asymmetry of the subbands rapidly decreases as $S$ increases.
Inside each subband, near its edge at $E_i$,
$D(E) \sim |E-E_i|^{-1/2}$ for $L \rightarrow \infty$.

\item Schematic structure of the low-lying states
for spin $S=\frac{1}{2}$,
$h=0$ and $\epsilon\neq 0$, for A) $L$ even and B) $L$ odd.

\item
Spectrum of ${\cal H}_{\rm yz}$ for spin
$S=\frac{1}{2}$ and $\epsilon=0.25$
as a function of $h$ for $L=10$ sites. The two vertical lines indicate
the location of the transitions between the O/F
phases and the F/D phases,
respectively, for $L\rightarrow\infty$.

\end{list}
\newpage
\centerline{\psfig{figure=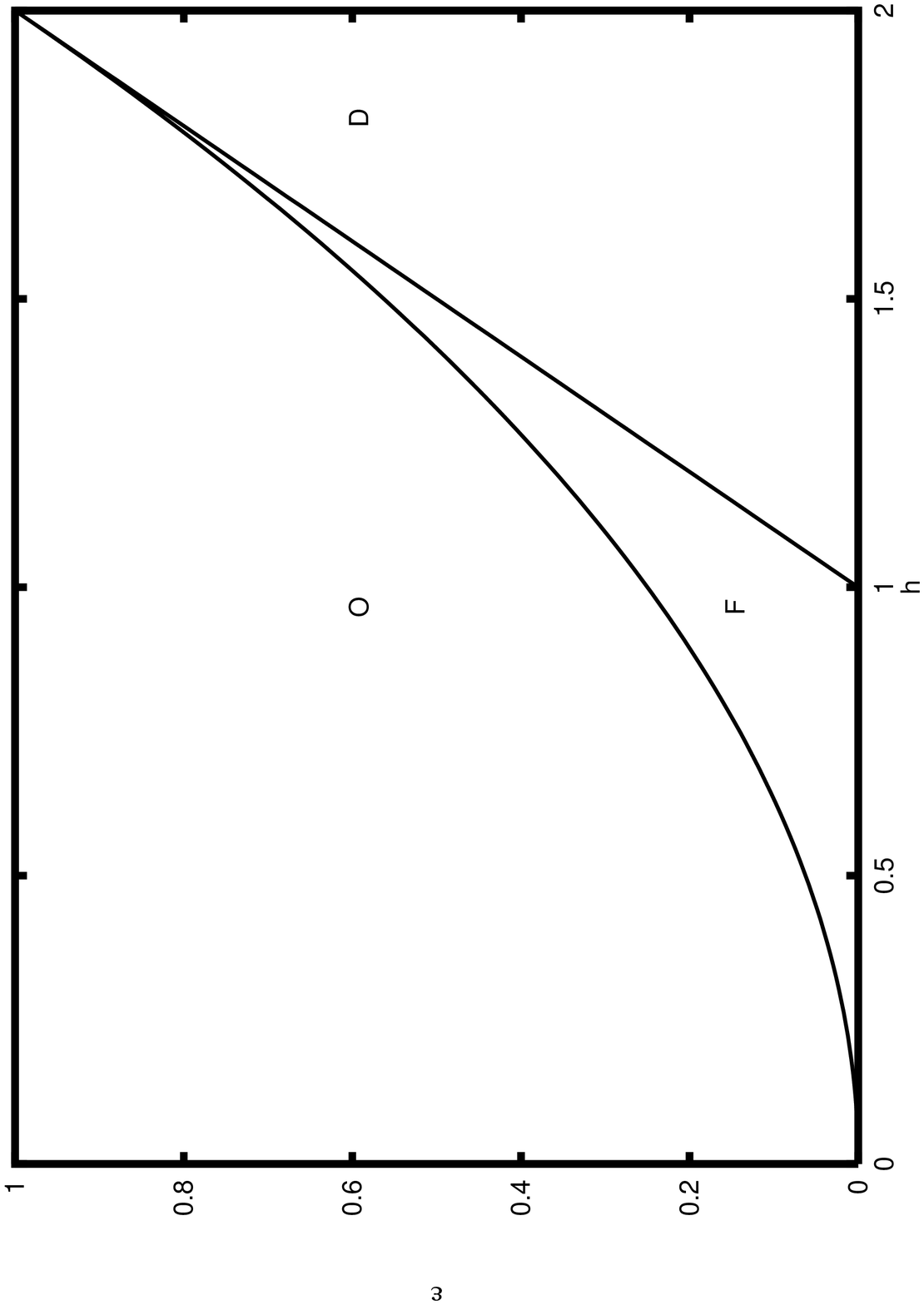,height=21cm}}
\newpage
\centerline{\psfig{figure=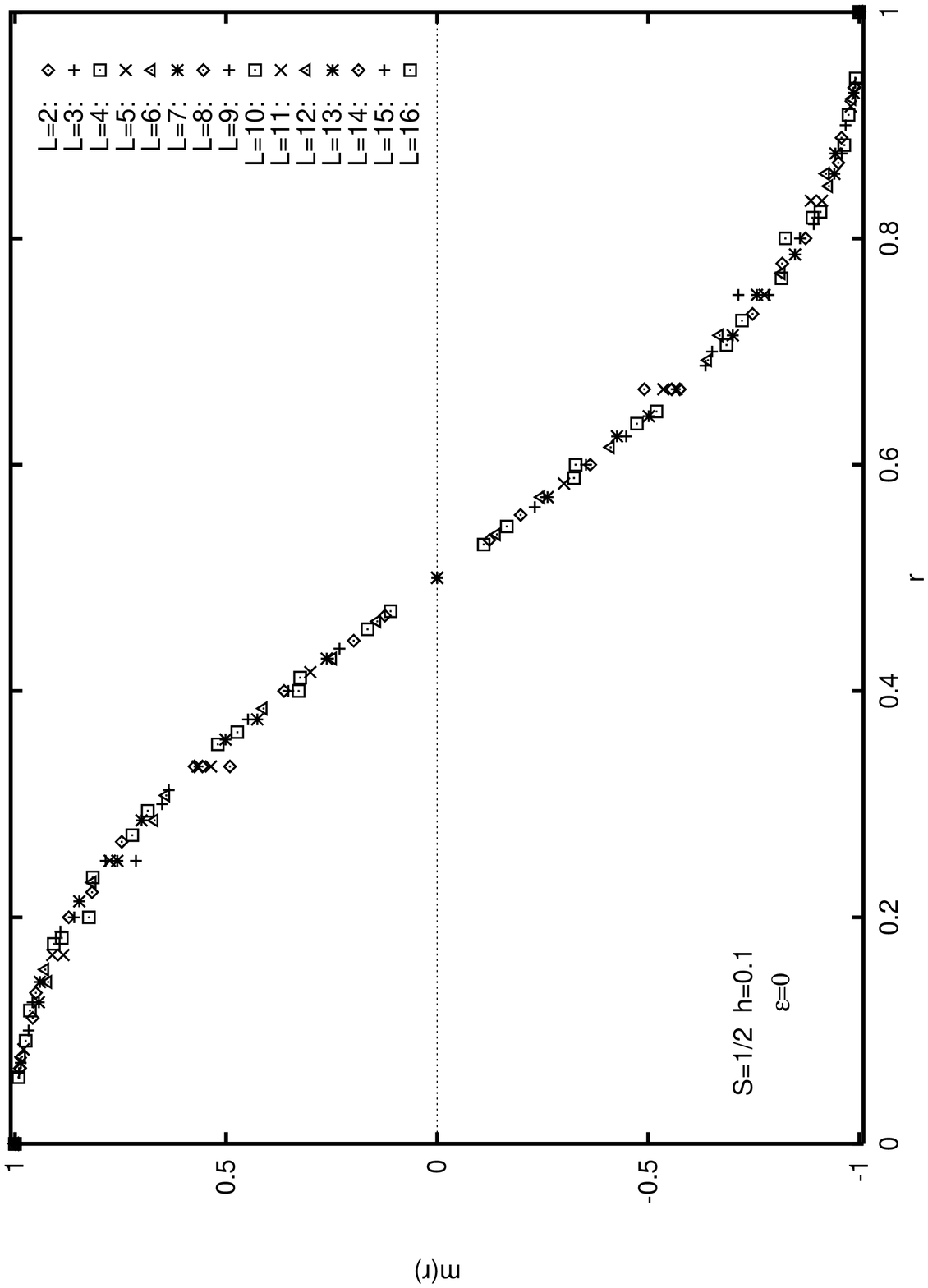,height=21cm}}
\newpage
\centerline{\psfig{figure=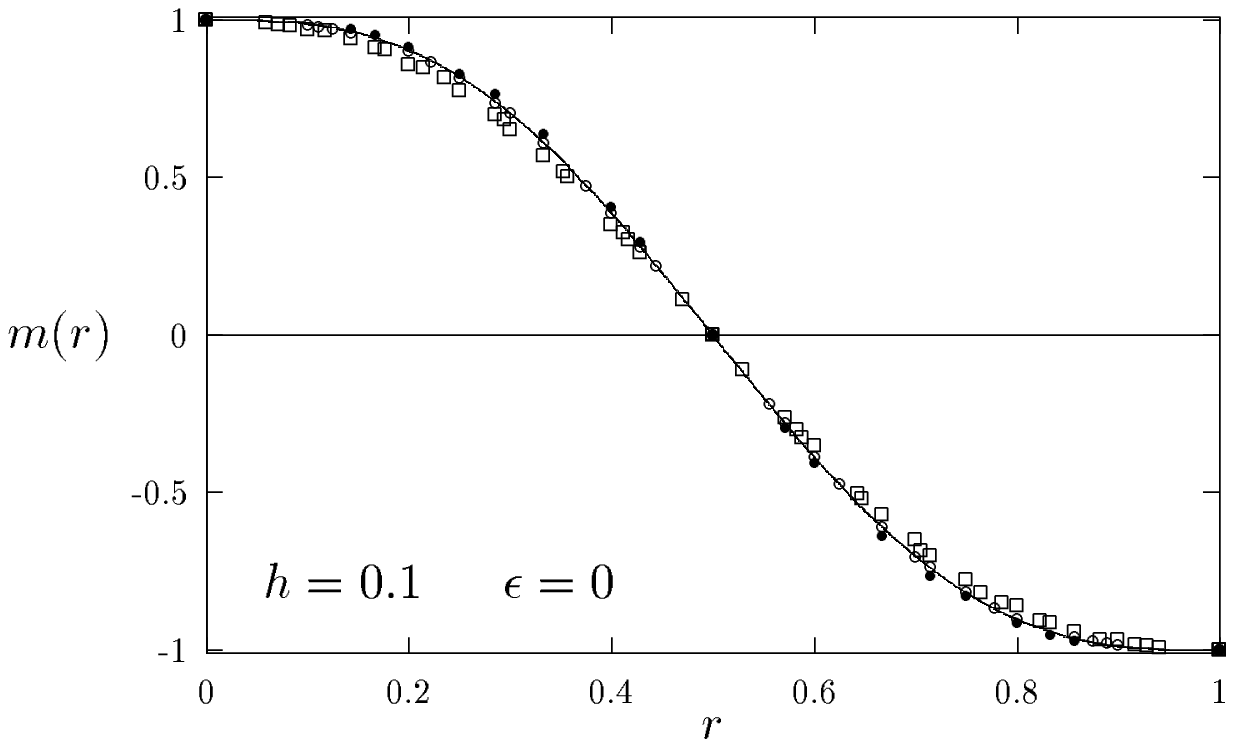}}
\newpage
\centerline{\psfig{figure=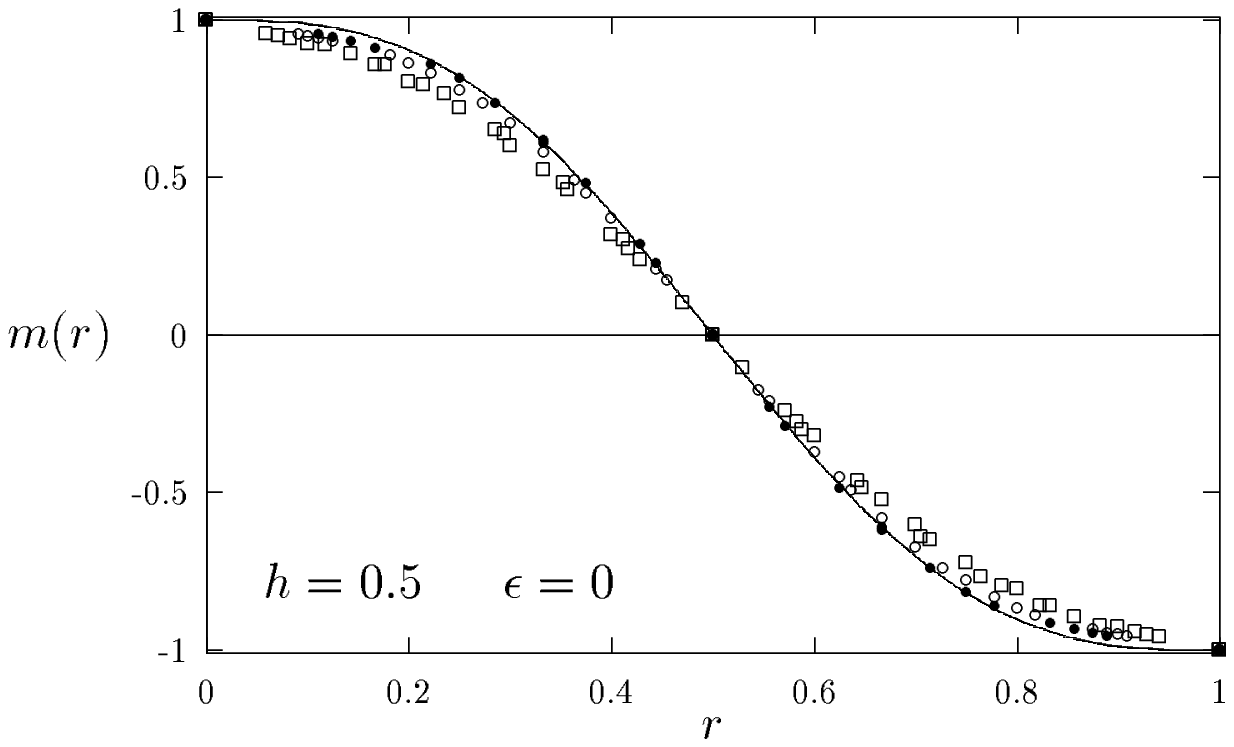}}
\newpage
\centerline{\psfig{figure=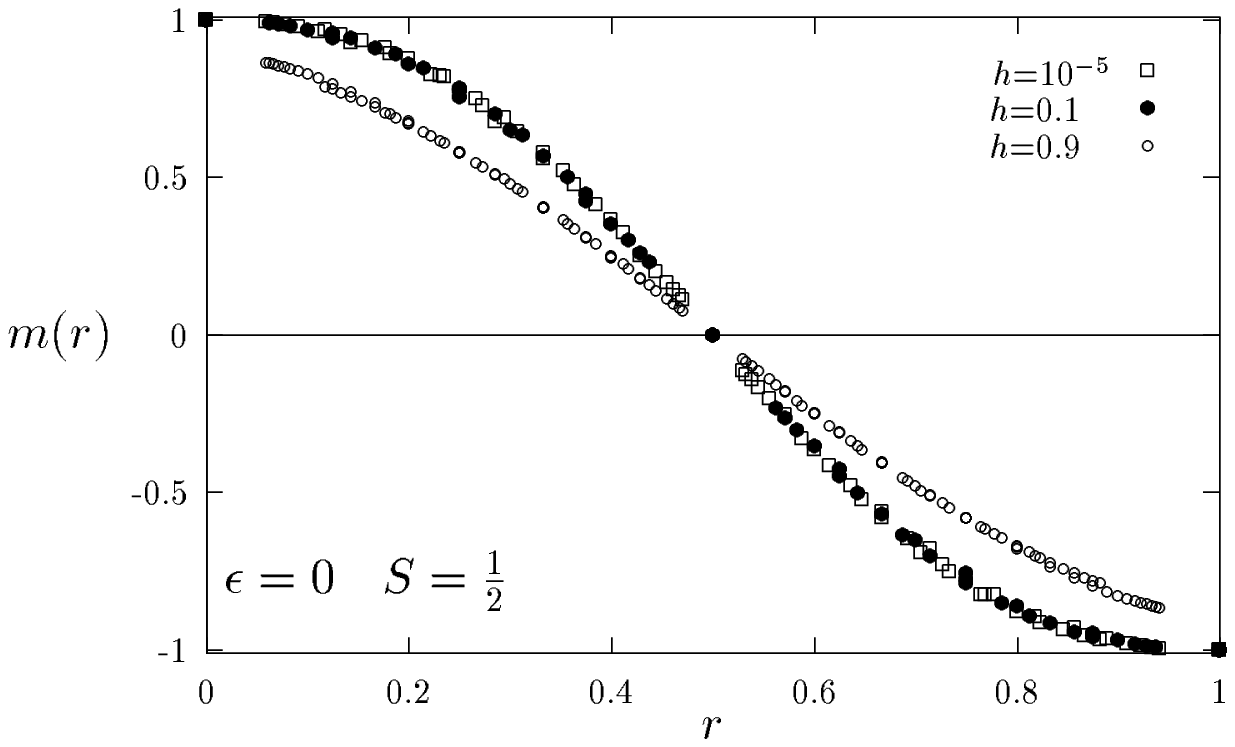}}
\newpage
\centerline{\psfig{figure=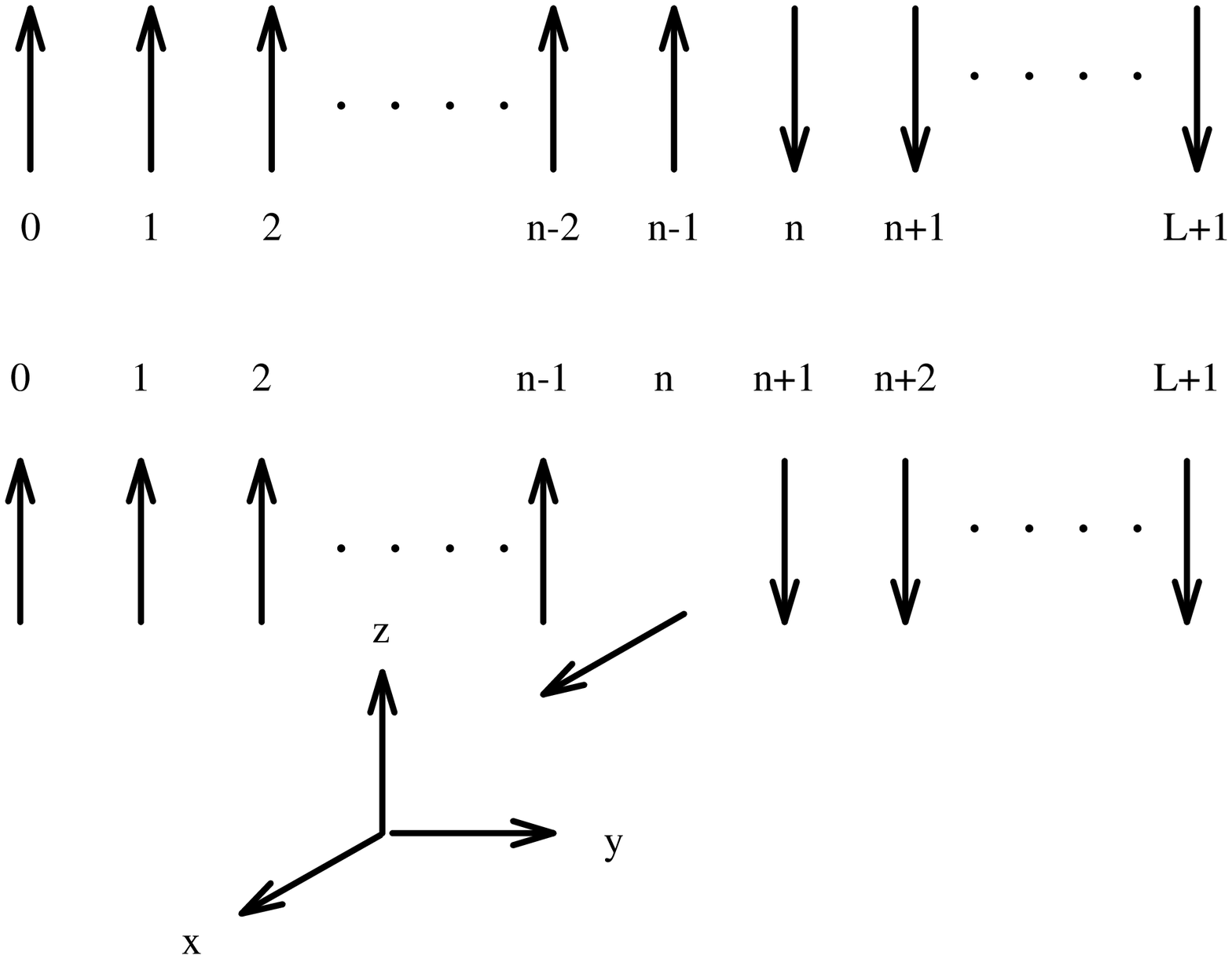}}
\newpage
\centerline{\psfig{figure=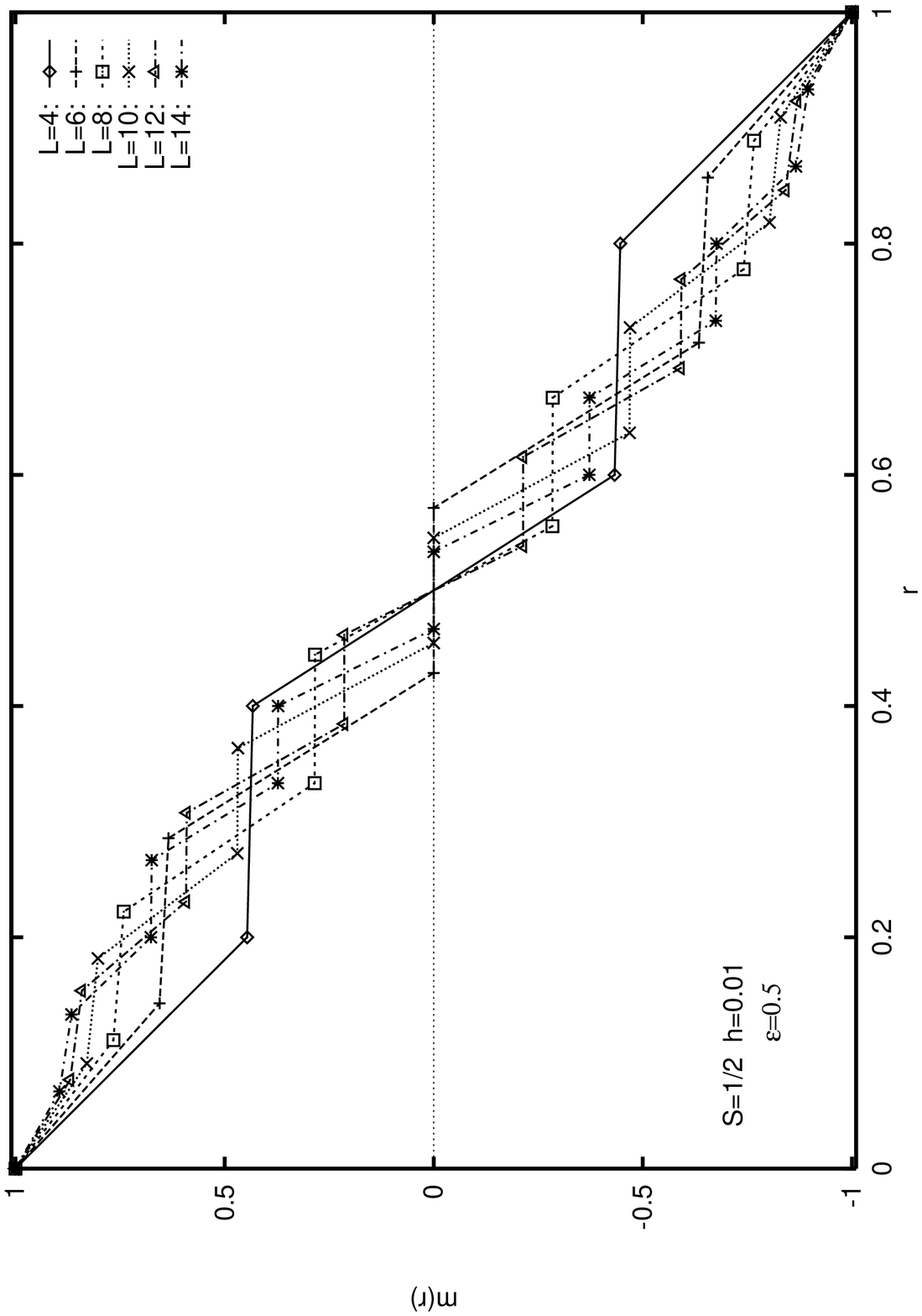,height=21cm}}
\newpage
\centerline{\psfig{figure=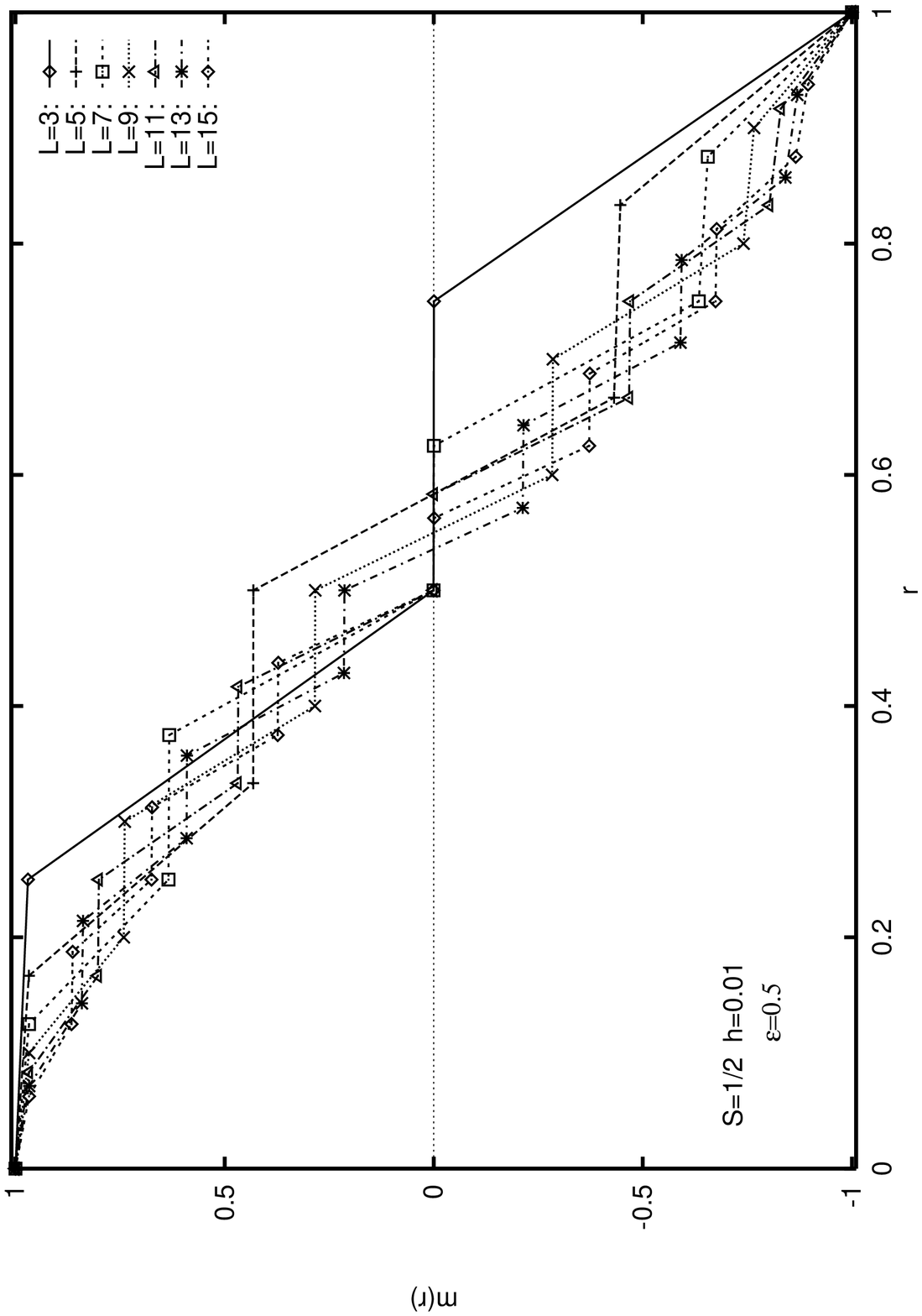,height=21cm}}
\vspace{3cm}
\newpage
\centerline{\psfig{figure=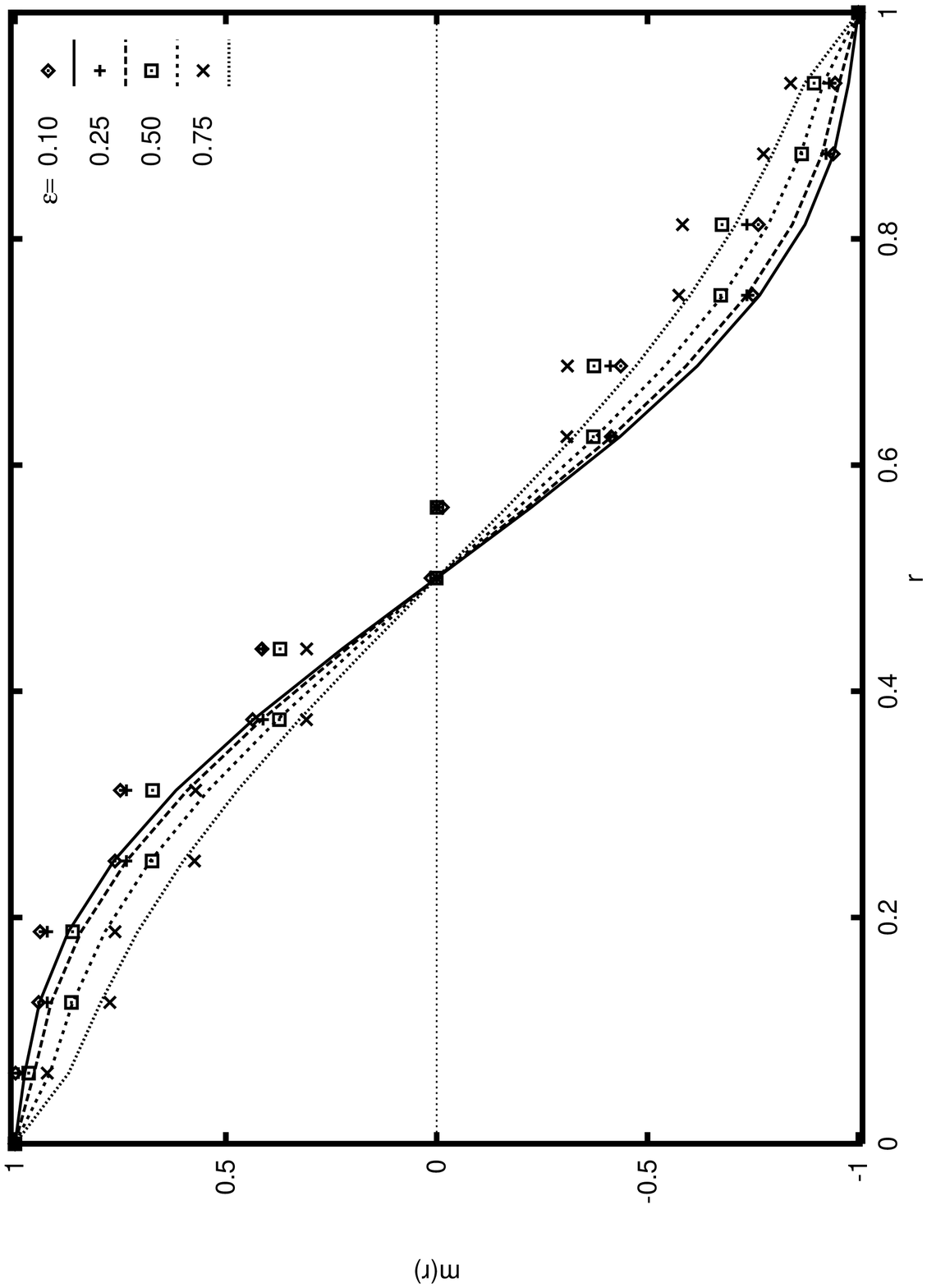,height=21cm}}
\newpage
\centerline{\psfig{figure=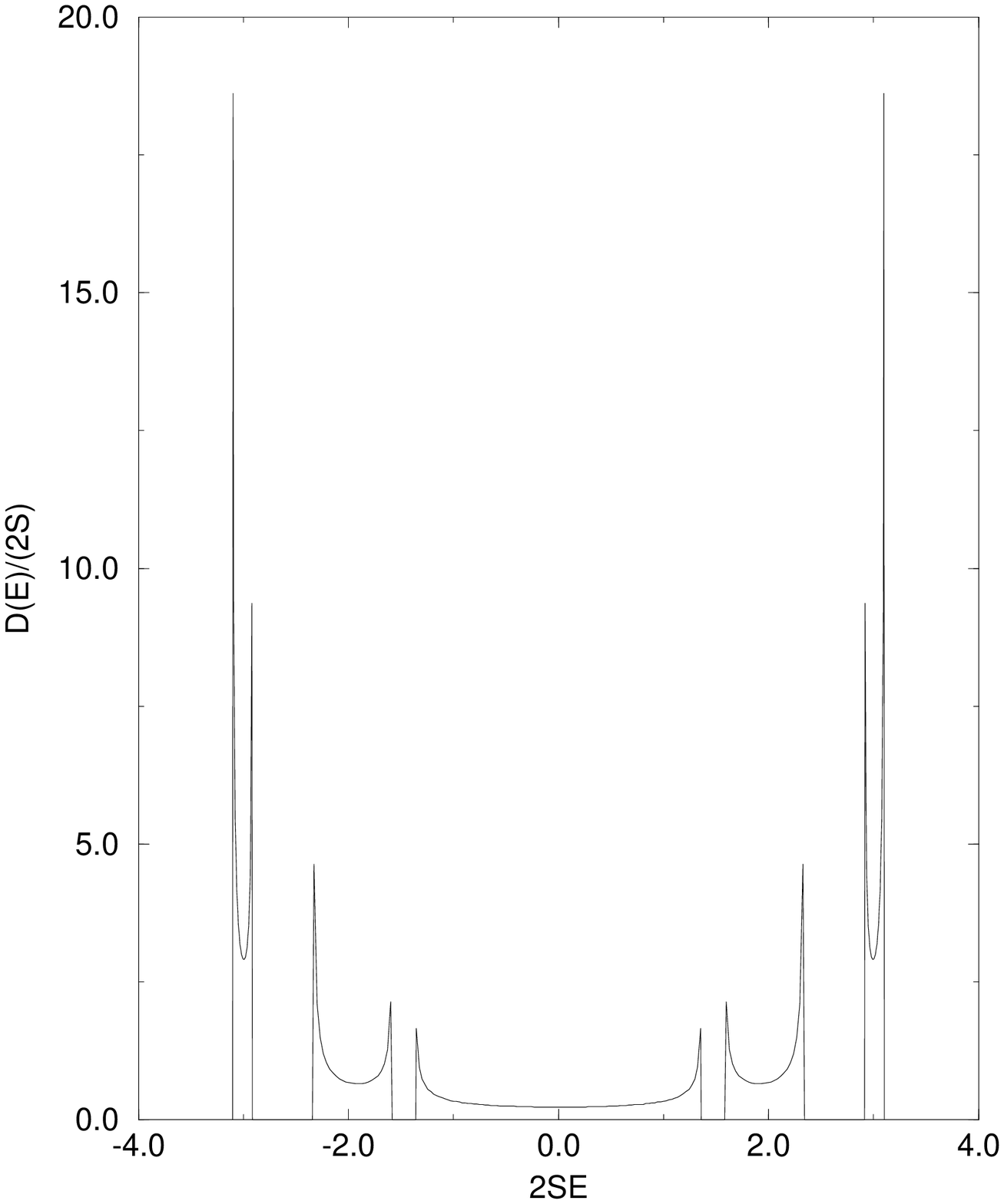}}
\newpage
\centerline{\psfig{figure=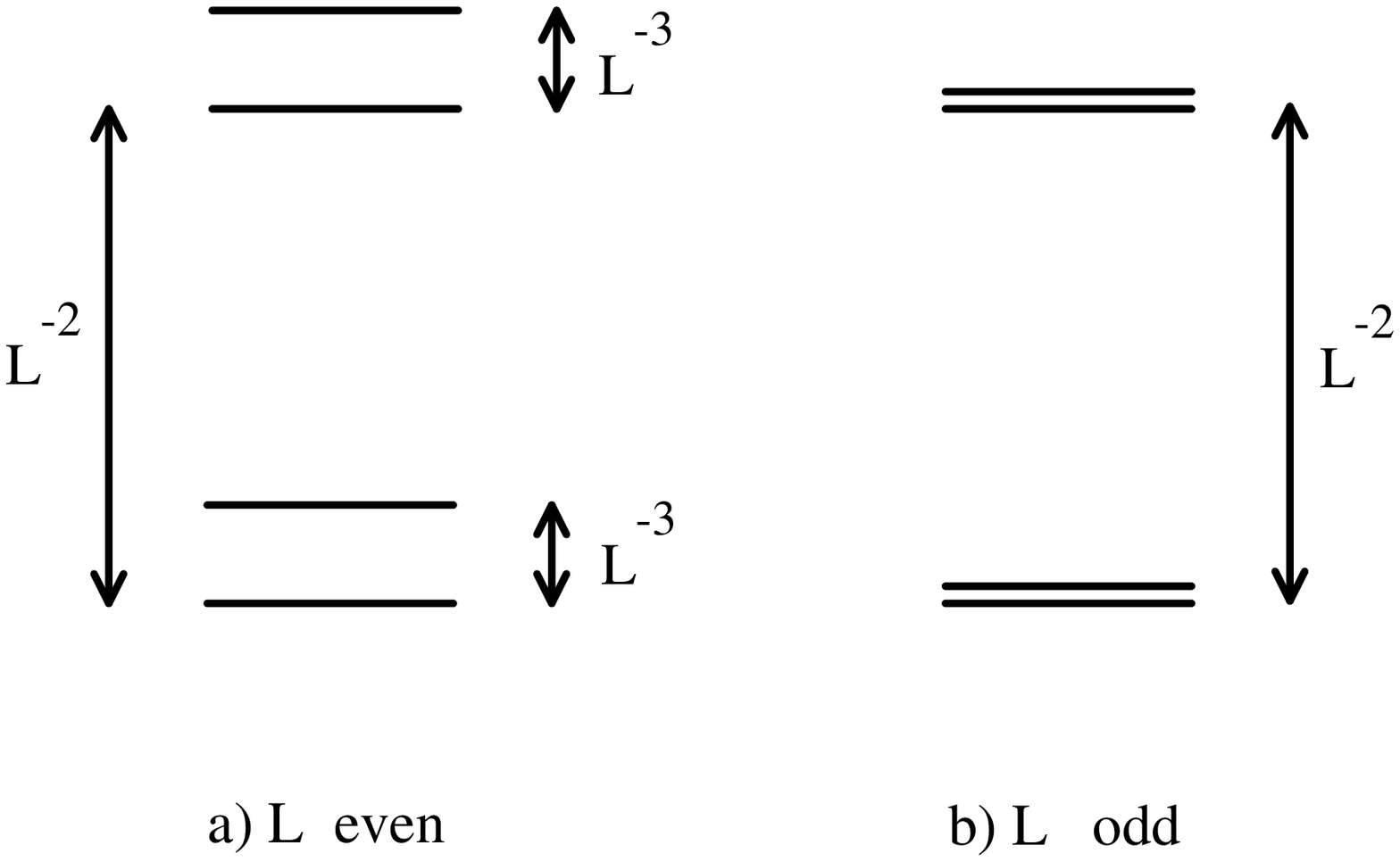}}
\newpage
\centerline{\psfig{figure=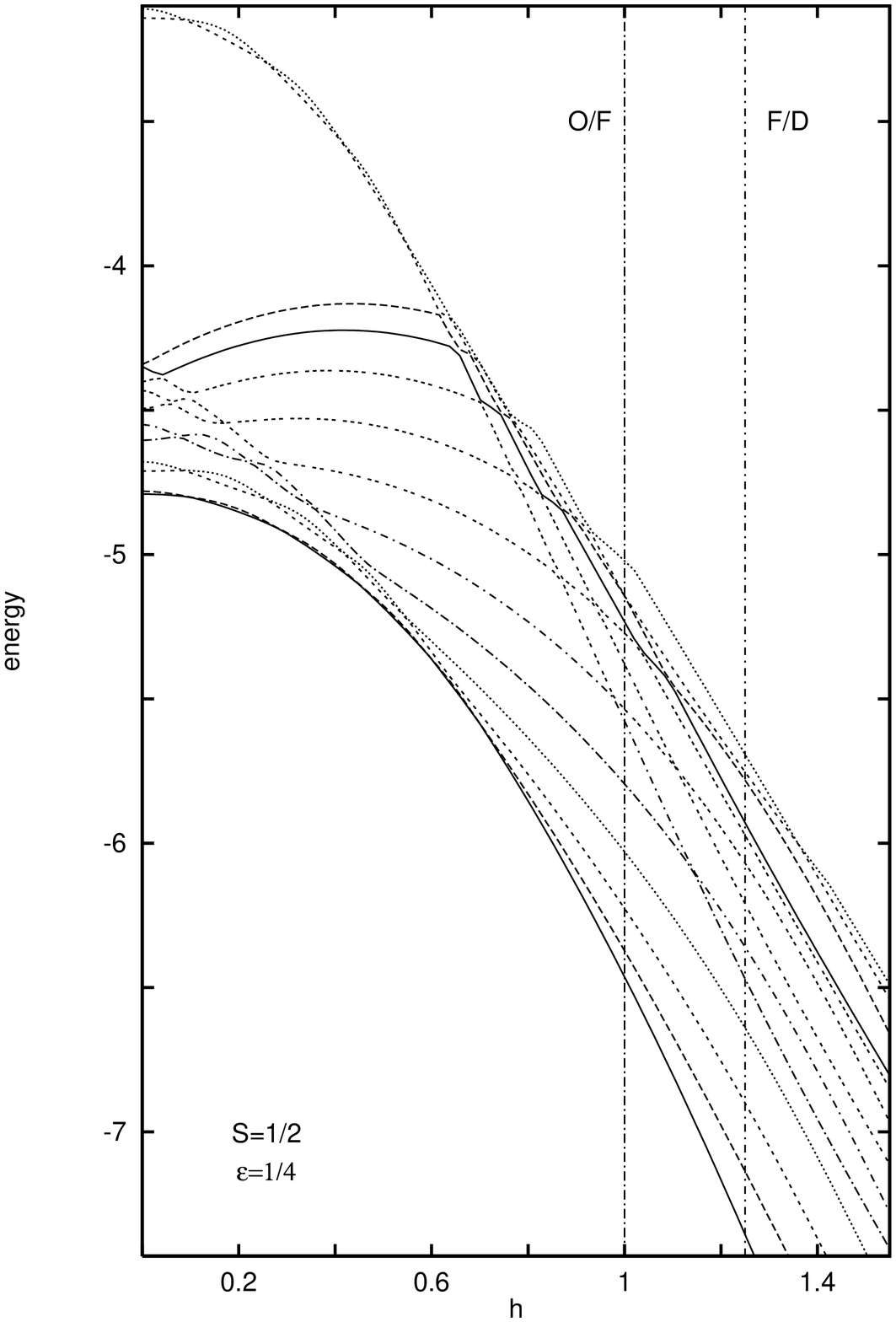,height=21cm}}


\begin{references}
\bibitem{BAND}
This argument is correct to leading order in $\tau$.  It
neglects overlap between ${\cal H}|n\rangle$ and other states
of ${\cal M}$, notably states similar to $|n \rangle$, but in
which $S_{nx}$ has the value $S-1$.  However the band of the
$S_x=S-1$ states is separated from the band of $S_x=S$ states
by an energy approximately equal to $H/(2S)$.  But since the relevant
matrix element is of order $\tau H << H$ and is off-diagonal, it
will enter at order $(\tau H)^2/[H/(2S)] \sim S\tau^2$.  Thus this
effect is negligible compared to the terms we do treat, at least
for large $S$.
\end{references}
\end{document}